\documentclass[a4paper,11pt]{article}
\usepackage{amsmath}
 
\usepackage{amsfonts}
\usepackage{tabularx,ragged2e,booktabs,caption}
\usepackage{setspace}
\usepackage{amssymb}
\usepackage{color}
\usepackage[round,authoryear]{natbib}
\usepackage{graphicx}
\usepackage{graphics}
\usepackage{lineno}
\usepackage{lipsum}
\usepackage[T1]{fontenc}
\usepackage[utf8]{inputenc}
\usepackage{authblk}
\usepackage{enumitem}
\usepackage{array}
\newcolumntype{L}[1]{>{\raggedright\arraybackslash}p{#1}}
\newcolumntype{C}[1]{>{\centering\arraybackslash}p{#1}}
\newcolumntype{R}[1]{>{\raggedleft\arraybackslash}p{#1}}
\usepackage{geometry}
\usepackage{booktabs}
\usepackage{tabularx, array}
\usepackage{dcolumn}

\begin{document}

\title{Asset Allocation Strategies Based on Penalized Quantile Regression}

\author[1]{G. Bonaccolto}
\author[2]{M. Caporin}
\author[3]{S. Paterlini}
\affil[1]{\small{Department of Statistical Sciences, University of Padova, via C. Battisti 241, 35121 Padova, Italy. $Email:\;bonaccolto@stat.unipd.it$}}
\affil[2]{Department of Economics and Management ``Marco Fanno'', via del Santo 33, 35123 Padova, Italy. $Email: massimiliano.caporin@unipd.it$}
\affil[3]{Department of Finance and Accounting, European Business School, Gustav-Stresemann-Ring 3, 65189 Wiesbaden, Germany. $Email:sandra.paterlini@ebs.edu$}
\date{}
\maketitle

\begin{abstract}
It is well known that quantile regression model minimizes the portfolio extreme risk, whenever the attention is placed on the estimation of the response variable left quantiles. We show that, by considering the entire conditional distribution of the dependent variable, it is possible to optimize different risk and  performance indicators. In particular, we introduce a risk-adjusted profitability measure, useful in evaluating financial portfolios under a pessimistic perspective, since the reward contribution is net of the most favorable outcomes. Moreover, as we consider large portfolios, we also cope with the dimensionality issue by introducing an $\ell_1$-norm penalty on the assets weights.   
\end{abstract}

\textbf{Keywords:} Quantile regression, $\ell_1$-norm penalty, pessimistic asset allocation.

\textbf{JEL codes:} C58, G10.

\section{Introduction}

Starting with the seminal contribution by \cite{Ma52} on the mean-variance portfolio theory, portfolio estimation and asset selection got increasing attention from both a practitioner's and a research view point. In the finance industry, asset allocation and security selection have a central role in designing portfolio strategies for both private and institutional investors. Differently, the academia focused on developments of the Markowitz approach over different research lines: linking it to market equilibrium as done by  \cite{Sharpe64}, \cite{Lintner65a}, \cite{Lintner65b}, and \cite{Mossin66}; modifying the objective function both when it is set as an utility function or when it takes the form of a performance measure \citep{AleBap2002,FaFeRoThTi2008}; developing tools for the estimation and the forecasting of the Markowitz model inputs, with great emphasis on return and risk.

Among the various methodological advancements, we focus on those associated with variations of the objective function or, more generally, based on alternative representations of the asset allocation problem. Some of the various asset allocation approaches proposed in the last decades share a common feature: they have a companion representation in the form of regression models where the coefficients correspond or are linked to the assets weights in a portfolio. Two examples are given by the estimation of efficient portfolio weights by means of linear regression of a constant on asset excess returns \citep{BJ99} and the estimation of the Global Minimum Variance Portfolio weights through the solution of a specific regression model, see e.g. \cite{FaZhYu12}.

In the previously cited cases, the portfolio variance plays a fundamental role. However, even if we agree with the relevance of variance (or volatility) for risk measurement and management, the financial literature now includes a large number of other indicators that might be more appropriate. As an example, for an investor whose preferences or attitudes to risk are summarized by an utility function where extreme risk is present, volatility might be replaced by  tail expectation. Similarly, we can easily identify reward measures and performance indicators differing from the simple average (or cumulated) return and from the Sharpe ratio. On the one side, we aim at keeping those elements in mind but, on the other side, we want to remain confined within the allocation approaches where weights can be associated with a linear model. In such a framework, it is possible to show that the adoption of non standard methods for the estimation of the linear model parameters (and portfolio weights) leads to solutions equivalent to the optimization of performance measures or risk measures differing from the Sharpe ratio and the volatility. Thus, moving away from the least square regression approach for estimating portfolio weights is equivalent to optimizing a non-standard objective function.

The leading example is given by  \cite{BaKoKo04}, which proposed a pessimistic asset allocation strategy relying on the quantile regression method introduced by \cite{KoBa78}. In particular, \cite{BaKoKo04} start from the linear model whose solution provides the global minimum variance portfolio weights. Then, they show that estimating by quantile regression methods a low quantile (the $\alpha$-quantile) of the response variable allows minimizing a measure of the portfolio extreme risk that they call $\alpha$-risk. Therefore, a change in the estimation approach allows moving from Global Minimum Variance portfolio to the minimum $\alpha$-risk portfolio.

Variants of the $\alpha$-risk are known under a variety of names, such as ``Expected Shortfall'' \citep{AcTa02}, ``Conditional Value-at-Risk'' \citep{RoUr00}, and ``Tail Conditional Expectation'' \citep{ArDeEbHe99}. Consequently, the pessimistic asset allocation strategy of \cite{BaKoKo04} corresponds to an \textit{extreme} risk minimization approach.

The work by \cite{BaKoKo04} also represents the starting point of our contributions. Building on quantile regression methods, we aim at introducing innovative asset allocation strategies coherent with the maximization of a risk-reward trade-off. Moreover, we combine quantile regression with regularization methods, such as LASSO \citep{Ti96}, in order to cope with the problematic issues arising form the large portfolios dimensionality.  Our contributions provide an answer to specific research questions, with a potential application within the financial industry. 

The first research question originates from a limitation of the pessimistic asset allocation approach of \cite{BaKoKo04}, which is a risk-minimization-driven strategy. Is it possible to maintain the focus on the $\alpha$-risk, as in \cite{BaKoKo04}, and at the same time maximize a performance measure, thus taking into account also rewards?
Our first contribution consists in showing that quantile regression models can be used not only to build financial portfolios with minimum extreme risk, as well known in the financial econometrics literature, but also to optimize other risk and performance measures by exploiting the information contained in the entire support of the response variable conditional distribution. We focus on linear model representation whose coefficients are associated with the Global Minimum Variance portfolio weights, as in \cite{BaKoKo04}, where quantile regression estimation of the linear model coefficients leads to the minimum $\alpha$-risk portfolio. We generalize the results in two ways. First, showing that, under reasonable assumptions, at the median level, the quantile regression solution of the linear model corresponds to the minimization of the mean absolute deviation of portfolio returns. Secondly, at high quantiles levels the quantile regression solution provides portfolio weights with an outstanding performance in terms of profitability and risk-adjusted returns. Such a solution corresponds to the maximization of a specific reward measure, which is given as the conditional expected return net of the most favorable outcomes; hence, it is a pessimistic allocation, as in \cite{BaKoKo04}, but with focus on the right tail rather than the left one. As a by-product, we introduce a performance measure, completely new in the literature; it is a risk-adjusted ratio which 
quantifies the magnitude of all the negative returns balanced by a part of positive results, net of the most favorable ones.

The second research question stems from empirical evidence and practitioners' needs. Financial portfolios are frequently characterized by a large cross-sectional dimension, i.e. they (potentially) include a large number of assets. Assuming we are interested in maintaining a pessimistic asset allocation strategy, possibly coherent with the investor preferences, we face a clear trade-off: on the one side the large cross-sectional dimension allows taking advantage of the diversification benefits which are anyway relevant even within a pessimistic allocation approach; on the other side, the number of parameters to estimate by a quantile regression approach quickly increases as the portfolio dimension grows. As a result, the accumulation of estimation errors becomes a problem that must be addressed. The question is as follows: can we control estimation error by maintaining the focus on the pessimistic asset allocation approach? Providing a possible solution, we impose a penalty on the $\ell_1$-norm of the quantile regression coefficients along the line of the Least Absolute Shrinkage and Selection Operator (LASSO), introduced by \cite{Ti96} in a standard linear regression framework. Recent studies show that applications of the LASSO to the mean-variance portfolio framework provide benefits in terms of sparsity of the portfolio (indirectly associated with diversification/concentration and turnover) and good out-of-sample properties, see e.g. \cite{BrDa09}, \cite{DMGaNoUp09}, \cite{FaZhYu12}, \cite{YeYe14} and \cite{FaPaWi14}. In the statistical literature, the $\ell_1$ penalty became a widely used tool 
not only in linear regression, but also for quantile regression models, see, e.g. \cite{Ko05}, \cite{BeCh11}, \cite{LiZh08}, while applications in asset allocation are still scarce. \cite{HaNa14} used penalized quantile regression as a security selection tool, in an index tracking framework, whereas the portfolios weights are  estimated by optimizing as objective function the Cornish-Fisher Value-at-Risk. Differently, in the approach we introduce, the penalized quantile regression model automatically selects and estimates the relevant assets weights in a single step. To the best of our knowledge, such an approach has never been investigated in the financial econometrics literature. We therefore suggest to estimate the pessimistic asset allocation strategy of \cite{BaKoKo04} and the pessimistic allocation here introduced, by penalizing the $\ell_1$-norm of the assets weights.  

We evaluate the two methodological contributions previously outlined with an extensive empirical analysis, in which we compare the performance of the asset allocation strategies built from the quantile regression models, at different quantiles levels, and the ordinary least squares approach. Differently from \cite{BaKoKo04}, we use both simulated as well as real-world data. Moreover, we analyze both the in-sample and the out-of-sample performances by implementing a rolling window procedure. Finally, we focus on portfolios with a reasonably large cross-sectional dimension, being thus close to the real needs of the financial industry.
The in-sample results, on both real-world and simulated data, show that each strategy performs consistently to the expectations, optimizing the respective objective functions, $\alpha$-risk, mean absolute deviation, or the upper-tail-based reward measure. Indeed, quantile regression applied at low probability levels outperforms the other strategies in terms of extreme risk. Least squares and median regression  models turn out to be the best strategies in terms of volatility, as the former minimizes the portfolio variance whereas the latter minimizes the mean absolute deviation of portfolios returns. Finally, quantile regression at high probability levels provides the best results in terms of profitability and risk-adjusted return. Out-of-sample results show that the quantile regression models maintain their in-sample properties but only at high probability levels. Despite such a result might be interesting from a practitioner's point of view, it is quite surprising from a methodological perspective. We studied this phenomenon providing an explanation associated with the role of the model intercept. Finally, we highlight the critical importance of regularizing the quantile regression problem to improve the out-of-sample performance of portfolios characterized by large cross-sectional dimension.

The work is structured as follows. In Section 2 we show
how the quantile regression model can be used to build asset allocation strategies optimizing different objective functions at different quantiles levels. In Section 3 we provide the details about the datasets, the performance indicators, the rolling window procedure and the empirical set-up. In Section 4 we discuss the empirical results, and Section 5 concludes.

\section{Asset allocation based on quantile regression} \label{sec:method_all}

\subsection{Portfolio performance as function of quantiles levels} \label{sec:method1}

Several asset allocation strategies estimate portfolio weights by optimizing a criterion function. The latter could either take the form of an utility function, of a risk measure, of a performance measure or a combination of different measures. A subset of those asset allocation approaches have a companion representation in the form of a regression model where the estimated coefficients correspond to the portfolio weights. The leading example is the Global Minimum Variance Portfolio ($GMVP$), whose composition is the solution of the ordinary least squares regression model.  

In the case of a financial portfolio consisting of $n$ stocks, let $\textbf{R}=[R_1,...,R_n]$ be the row vector of the assets returns\footnote{To simplify the notation we suppress the dependence of returns on time.}, with covariance matrix $\mathbb{V}\left[\textbf{R}\right]=\pmb{\Sigma}$, whereas the  row vector of weights is denoted by $\textbf{w}=[w_1,...,w_n]$; given the row vector $\mathbf{1}$, with length $n$ and elements equal to 1, $\mathbf{1} \textbf{w}'=1$ in order to satisfy the budget constraint. The portfolio return is then $R_p=\textbf{R} \textbf{w}'$, but we can also use a companion representation to include the budget constraint. First, we set $R^*_i=R_n-R_i$, for $i=1,...,n-1$, and then use these deviations for the computation of the portfolio return which becomes $R_p=R_n-w_1 R^*_1-...-w_{n-1}R^*_{n-1}$, where the $n-$th asset weight ensures the weights sum at 1. By starting from such representation, it is possible to show that the minimization of the portfolio variance can be expressed as
\begin{eqnarray} \label{fanminvareq}
\operatorname*{min}_{ \textbf{w}  \in \mathbb{R}^{n}} \textbf{w}\pmb{\Sigma}\textbf{w}' =\operatorname*{min}_{ (\textbf{w}_{-n},\xi)  \in \mathbb{R}^{n}} \mathbb{E} \left[ \textbf{R} \textbf{w}'-\xi \right]^2 = \operatorname*{min}_{(\textbf{w}_{-n},\xi) \in \mathbb{R}^n} \mathbb{E} \left[R_n-w_1 R^*_1-...-w_{n-1}R^*_{n-1}- \xi \right]^2,
\end{eqnarray}
where $\xi$ is the intercept of the linear regression model, 
$\textbf{w}_{-n}$ denotes the weights vector excluding $w_n$ and $w_n=1-\sum_{i=1}^{n-1}w_i$, in order to satisfy the budget constraint. 

In Equation (\ref{fanminvareq}), the portfolio variance, $\textbf{w}\pmb{\Sigma}\textbf{w}'$, is rewritten as $\mathbb{E} [R_n-w_1 R^*_1-...-w_{n-1}R^*_{n-1}$
$-\xi]^2$. The latter, corresponds to the variance of the errors for the linear regression of asset $n$ returns, $R_n$, with respect to $R_i^*$. Therefore, it is possible to minimize $\textbf{w}\pmb{\Sigma}\textbf{w}'$ by minimizing the sum of squared errors of a linear regression model, with response variable $R_n$ and covariates $R^*_1,...,R^*_{n-1}$. Thus, estimating the coefficients $w_1,...,w_{n-1}$, along with the intercept $\xi$, is equivalent to finding the $GMVP$ weights \citep{FaZhYu12}. In Model (\ref{fanminvareq}), the response variable equals the $n$-th asset returns, $R_n$. However, the result does not depend on the choice of the numeraire asset.

Despite the portfolio variance is a relevant risk measure, the financial literature presents a large number of other indicators to be considered for profitability and risk analyses. If we move away from the standard linear regression framework above mentioned, it is possible to estimate portfolios compositions by optimizing alternative performance measures. For instance, \cite{BaKoKo04} proposed a pessimistic asset allocation strategy that relies on 
quantile regression in order to minimize a risk measure: the so-called $\alpha-$risk.

By starting from a monotone increasing utility function $u(\cdot)$, that transforms monetary outcomes into utility terms, \cite{BaKoKo04} define the expected utility associated with $R_p$ as
\begin{equation} \label{exput}
\mathbb{E}[u(R_p)]=\int_{-\infty}^{\infty} u(r_p)dF_{R_p}(r_p)=\int_{0}^{1} u \left( F_{R_p}^{-1}(\vartheta) \right)d\vartheta,
\end{equation}
where $F_{R_p}(r_p)$ denotes the distribution function of $R_p$ evaluated at $r_p$, with density $f_{R_p}(r_p)$, whereas $\vartheta$ is the quantile index such that $\vartheta \in \mathcal{U}$, with 
$\mathcal{U} \subset (0,1)$.  

Such a framework is directly linked with the Choquet expected utility \citep{Sc89}, defined as
\begin{equation} \label{ChoExpUt}
\mathbb{E}_\nu[u(R_p)]=\int_{0}^{1} u \left( F_{R_p}^{-1}(\vartheta) \right)d \nu (\vartheta),
\end{equation}
where $\nu(.)$ is a distortion function which modifies the original probability assessment. In particular, $\nu(.)$ allows inflating the likelihood associated with the least favorable (i.e. $\nu(.)$ is concave) or the most favorable outcomes (i.e. $\nu(.)$ is convex). \cite{BaKoKo04} propose to use the function $\nu_\alpha(\vartheta)=\operatorname*{min}\{\vartheta/\alpha,1\} $, where $\alpha$ is a very low probability level, e.g. $\alpha=\{0.01,0.05,0.1\}$, associated with the (negative) returns in the left tail of $f_{R_p}(r_p)$. Then, (\ref{ChoExpUt}) can be rewritten as
\begin{equation} \label{ExpChoq2}
\mathbb{E}_{\nu_\alpha}[u(R_p)]=\alpha^{-1}\int_{0}^{\alpha} u \left( F_{R_p}^{-1}(\vartheta) \right)d \vartheta,
\end{equation} 
where $\mathbb{E}_{\nu_\alpha}[u(R_p)]$ implies pessimism since it inflates the likelihood associated with the $\alpha$ least favorable outcomes, whereas the remaining $1-\alpha$ proportion is entirely deflated. 

Equation (\ref{ExpChoq2}) is directly linked to a measure of extreme risk, $\varrho_{\nu_\alpha}(R_p)$, defined by \cite{BaKoKo04} as $\alpha$-risk: 
\begin{equation} \label{alpharisk}
\varrho_{\nu_\alpha}(R_p)=-\int_{0}^{1}F_{R_p}^{-1}(\vartheta)d\nu(\vartheta)=-\alpha^{-1}\int_{0}^{\alpha}F_{R_p}^{-1}(\vartheta)d\vartheta.
\end{equation}
Such a quantity is a coherent measure of risk according to the definition of \cite{ArDeEbHe99}. Many variants of $\varrho_{\nu_\alpha}(R_p)$ have been discussed in the financial literature, under a variety of names: Expected Shortfall \citep{AcTa02}, Conditional Value-at-Risk \citep{RoUr00}, and Tail Conditional Expectation \citep{ArDeEbHe99}.\footnote{Although $\varrho_{\nu_\alpha}(R_p)$ could be denominated in different ways, throughout the paper we refer to the (\ref{alpharisk}) as $\alpha$-risk.}

Notably, (\ref{alpharisk}) might be taken as the target risk measure for portfolio allocation, see e.g. \cite{BaSh01}, \cite{KrPaUr02}, \cite{CiKoMe07}, \cite{MaOgSp07}. In such a case,  $\varrho_{\nu_\alpha}(R_p)$ can be minimized by resorting to the quantile regression method, as suggested by \cite{BaKoKo04}, in a framework similar to the estimation of the $GMVP$ weights in (\ref{fanminvareq}), where $R_n$ is the response variable, whereas $R^*_1,...,R^*_{n-1}$ are the covariates. Within a quantile regression framework, the conditional $\vartheta$-th quantile of $R_n$ is estimated by minimizing the expected value of the asymmetric loss function:

\begin{equation} \label{asylossfunctionqr}
\rho_\vartheta(\epsilon)=\epsilon\left[\vartheta-I(\epsilon<0)\right],
\end{equation}
where $\epsilon=R_n-\xi(\vartheta)-w_1(\vartheta)R^*_1-...-w_{n-1}(\vartheta)R^*_{n-1}$, $\xi(\vartheta)$ is the model intercept, and  $I(\cdot)$ denotes the indicator function taking value 1 if the condition in $(\cdot)$ is true, 0 otherwise. 

The estimated $\vartheta$-th conditional quantile of $R_n$ is equal to $\widehat{\xi}(\vartheta)+\widehat{w}_1(\vartheta)R^*_1+...+\widehat{w}_{n-1}(\vartheta)R^*_{n-1}$, where $\left[ \widehat{\xi}(\vartheta),\widehat{w}_1(\vartheta),...,\widehat{w}_{n-1}(\vartheta)\right]$ is the coefficients vector minimizing (\ref{asylossfunctionqr}), at a specific quantile level $\vartheta$ \citep{KoBa78}. In the case in which $\vartheta=\alpha$, \cite{BaKoKo04} showed that 
\begin{equation} \label{ffassdsafa}
\operatorname*{min}_{(\xi(\alpha),\textbf{w}_{-n}(\alpha)) \in \mathbb{R}^n}  \mathbb{E}[ \rho_\alpha(\epsilon) ]=\alpha \left( \mu_p + \varrho_{\nu_\alpha}(R_p)\right),
\end{equation}
where $\textbf{w}_{-n}(\alpha)=[w_1(\alpha),...,w_{n-1}(\alpha)]$, $\mu_p=\mathbb{E}[R_p]$ and $\varrho_{\nu_\alpha}(R_p)$ as in (\ref{alpharisk}).

Let $r_{n,t}$ and $r^*_{i,t}$ be, respectively, the observed values of $R_n$ and $R^*_i$, for $i=1,...,n-1$, at time $t$, then, from (\ref{ffassdsafa}), the quantile regression model
\begin{equation} \label{quantmodalprisk}
\operatorname*{min}_{(\xi(\alpha),\textbf{w}_{-n}(\alpha)) \in \mathbb{R}^n} \sum_{t=1}^{T} \rho_\alpha \left(r_{n,t}-w_1(\alpha) r^*_{1,t}-...-w_{n-1}(\alpha)r^*_{n-1,t} -\xi(\alpha) \right) \;\; s.t.\; \mu_p=c
\end{equation}
allows minimizing the empirical $\alpha$-risk of a financial portfolio, with the constraints that the expected portfolio return is equal to a target $c$ and that the sum of the assets weights is equal to 1. Similarly to Model (\ref{fanminvareq}), $\left[\widehat{w}_1(\alpha),...,\widehat{w}_{n-1}(\alpha)\right]$, the estimated coefficients vector of the covariates $R_1^*,...,R_{n-1}^*$ in the quantile regression model, is then the weights vector of $R_1,...,R_{n-1}$ for the portfolio with minimum $\alpha$-risk. Note that the weight of the $n$-th asset is then derived from the budget constraint: $w_n(\alpha)=1-\sum_{i=1}^{n-1}w_i(\alpha)$. In this formulation, the assets weights do not change if we change the numeraire. As the constraint 
$\mu_p=c$ in (\ref{quantmodalprisk}) requires the estimation of expected returns, which is known to be a challenging task due to 
large estimation errors, see e.g. \cite{Br93} and \cite{ChZi93}, we hereby choose to focus on  
\begin{equation} \label{quantmodalprisknocons}
\operatorname*{min}_{(\xi(\alpha),\textbf{w}_{-n}(\alpha)) \in \mathbb{R}^n} \sum_{t=1}^{T} \rho_\alpha \left(r_{n,t}-w_1(\alpha) r^*_{1,t}-...-w_{n-1}(\alpha)r^*_{n-1,t} -\xi(\alpha) \right),
\end{equation}
that is the minimization of the portfolio $\alpha$-risk, subject only to the budget constraint.  

As the portfolio performance does not depend just on extreme risk, but rather on the occurrence of returns over their entire density support, we generalize the approach of \cite{BaKoKo04} and allow the construction of portfolios  calibrated on different performance measures. 
First of all, we introduce an approach which makes use of two new performance measures, with potential application also in the performance evaluation area. The main idea stems from observing that (\ref{alpharisk}) could be associated with profitability, and no longer only with extreme risk, if we replace the low probability levels $\alpha$ by high probability levels. According to this intuition, we introduce two different indicators, as described next. If we denote the high probability values by $\psi$, e.g. $\psi=\{0.9,0.95,0.99\}$, the $\alpha$-risk in (\ref{alpharisk}) translates into
\begin{subequations}
\begin{equation}\setcounter{equation}{1}\label{PSI1}
\Psi_1(R_p,\psi)=-\psi^{-1}\int_{0}^{\psi} F_{R_p}^{-1}(\vartheta)d\vartheta. 
\end{equation}

Given that $-\Psi_1(R_p,\psi)=\mathbb{E}[R_p|R_p \leq F_{R_p}^{-1}(\psi)]$, the quantile regression model, applied at $\psi$, allows to minimize $\Psi_1(R_p,\psi)$ and, consequently, to maximize the
conditional portfolio expected return. Therefore, by minimizing $\Psi_1(R_p,\psi)$ an agent is taking a pessimistic asset allocation strategy, in the sense that such a choice leads to the maximization of the portfolio expected return net of the most favorable outcomes, since the interval $(\psi,1]$ is not included in the objective function.
Moreover, as $\lim_{\psi \rightarrow 1}-\Psi_1(R_p,\psi)=\int_{0}^{\psi} F_{R_p}^{-1}(\vartheta)d\vartheta$, it is possible to obtain benefits in terms of unconditional portfolio expected return, given that we maximize a quantity which approximates $\mathbb{E}[R_p]$. 
 
The second performance indicator we introduce is obtained by decomposing the integral in Equation (\ref{PSI1}). In particular, let $\bar{\vartheta}$ be the value of $\vartheta$ such that $F_{R_p}^{-1}(\bar{\vartheta})=0$, at which the integral $\int_{0}^{\bar{\vartheta}} F_{R_p}^{-1}(\vartheta)d\vartheta$ reaches its lowest value; for instance, $\bar{\vartheta}=0.5$ when the distribution is symmetric at $0$. Given $\bar{\vartheta}<\psi<1$, (\ref{PSI1}) could be rewritten as
\begin{equation}\setcounter{equation}{2}\label{PSI1b}
\Psi_1(R_p,\psi)=-\psi^{-1}\int_{0}^{\psi} F_{R_p}^{-1}(\vartheta)d\vartheta=-\psi^{-1} \left[\int_{0}^{\bar{\vartheta}} F_{R_p}^{-1}(\vartheta)d \vartheta +\int_{\bar{\vartheta}}^{\psi} F_{R_p}^{-1}(\vartheta)d \vartheta   \right], 
\end{equation}
\end{subequations} 
where $\int_{0}^{\bar{\vartheta}} F_{R_p}^{-1}(\vartheta)d \vartheta$ is computed from negative realizations and quantifies their magnitude. Differently, $\int_{\bar{\vartheta}}^{\psi} F_{R_p}^{-1}(\vartheta)d \vartheta$ quantifies the magnitude of a part of the positive outcomes, excluding the most favorable ones, given that the region beyond $\psi$ is not considered. The quantile regression model, applied at the $\psi$-th level, minimizes $\Psi_1(R_p,\psi)$ and, thus, also $-\varsigma=-\left( \int_{0}^{\bar{\vartheta}} F_{R_p}^{-1}(\vartheta)d \vartheta+\int_{\bar{\vartheta}}^{\psi} F_{R_p}^{-1}(\vartheta)d \vartheta\right)$. When $f_{R_p}(r_p)$ is characterized by a null or a negative skewness, $\varsigma$ is negative, whereas $\varsigma$ could be positive in the case of positive skewness. 
In the first case, $\varsigma$ could be seen as a net loss; differently, in the latter case, $\varsigma$ is a net profit. Therefore, quantile regression model leads to the minimization of a loss ($\varsigma<0$) or to the maximization of a profit ($\varsigma>0$), since, in  (\ref{PSI1b}), $\varsigma$ is multiplied by the constant $-\psi^{-1}<0$. In other words, the quantile regression model minimizes $|\varsigma|$, if $\varsigma<0$, or maximizes $|\varsigma|$, if $\varsigma>0$, with benefits in terms of the ratio:
\begin{equation} \label{PSI2}
\Psi_2(R_p,\psi)=\frac{\int_{\bar{\vartheta}}^{\psi} F_{R_p}^{-1}(\vartheta)d\vartheta}{\left| \int_{0}^{\bar{\vartheta}} F_{R_p}^{-1}(\vartheta)d\vartheta \right|}.
\end{equation}

Hence, the ratio $\Psi_2(R_p,\psi)$ is a risk-adjusted measure because it quantifies the magnitude of all the negative outcomes balanced by a part of positive results, net of the most favorable ones. Although high $\Psi_2(R_p,\psi)$ values correspond to low $\Psi_1(R_p,\psi)$ levels, when different strategies are compared, there are no guarantees that the strategy which minimizes $\Psi_1(R_p,\psi)$ is the one which maximizes $\Psi_2(R_p,\psi)$. In other words, the ranking of different strategies built on the sum between $\int_{\bar{\vartheta}}^{\psi} F_{R_p}^{-1}(\vartheta)d \vartheta$ and $\int_{0}^{\bar{\vartheta}} F_{R_p}^{-1}(\vartheta)d \vartheta$ may not coincide with the ranking built on the basis of their ratio. For example, suppose that for a certain strategy $A$, $\int_{0}^{\bar{\vartheta}} F_{R_p}^{-1}(\vartheta)d \vartheta=-34.04$ and  $\int_{\bar{\vartheta}}^{\psi} F_{R_p}^{-1}(\vartheta)d \vartheta=8.13$; differently, strategy $B$ returns $\int_{0}^{\bar{\vartheta}} F_{R_p}^{-1}(\vartheta)d \vartheta=-33.74$ and  $\int_{\bar{\vartheta}}^{\psi} F_{R_p}^{-1}(\vartheta)d \vartheta=7.95$. $B$ is better in terms of $\Psi_1(R_p,\psi)$, but $A$ outperforms $B$ in terms of $\Psi_2(R_p,\psi)$.

Therefore, beside their use within a quantile regression-based portfolio allocation strategy, the two indicators in (\ref{PSI1}) and (\ref{PSI2}) are also novel contributions for the performance measurement literature. We stress that while (\ref{PSI1}) resemble tail-based risk measures, and is not a proper absolute performance measure, see \cite{CaJaLiMa14},  indicator (\ref{PSI2}) is novel. It is interesting to notice that $\Psi_2(R_p,\psi)$ is related both to the \textit{Omega} measure  proposed by \cite{KeSh02} and to the modified Rachev Ratio \citep{OrRaStFaBi05}. Nevertheless, there are some important differences between these quantities.
The \textit{Omega} measure \citep{KeSh02} is defined as 
\begin{equation}
\Omega(R_p)=\frac{\int_{0}^{\infty}\left[1-F_{R_p}(r_p)dr_p\right]}{\int_{-\infty}^{0}F_{R_p}(r_p)dr_p}.
\end{equation} 
  
$\Psi_2(R_p,\psi)$ differs from Omega because the latter, compares the entire regions associated with negative and positive outcomes, respectively. Differently, (\ref{PSI2}) is more restrictive because its numerator takes into account just a part of positive outcomes, as long as $\psi<1$. 

The modified Rachev Ratio \citep{OrRaStFaBi05}, $MR(R_p,\alpha,\psi)$, equals 
\begin{eqnarray} \label{RachevRatio}
MR(R_p,\alpha,\psi)=\frac{-\alpha^{-1}\int_{0}^{\alpha}F^{-1}_{R_p}(\vartheta)d\vartheta}{(1-\psi)^{-1}\int_{\psi}^{1}F^{-1}_{R_p}(\vartheta)d\vartheta}.
\end{eqnarray}

In that case, the difference arises from the fact that (\ref{RachevRatio}) compares the extreme outcomes associated to the distribution tails, as typically $\alpha=\{0.01,0.05,0.1\}$ and $\psi=\{0.9,0.95,0.99\}$, thus fully neglecting the impact coming from the central part of the portfolio returns distribution.

In empirical application, we compute the sample counterparts of $\Psi_1(R_p,\psi)$ and $\Psi_2(R_p,\psi)$ as follows: 
\begin{eqnarray} \label{PSI1HAT}
\widehat{\Psi}_1(r_p,\psi)=-\frac{\sum_{t=1}^{T}r_{p,t} I \left( r_{p,t} \leq \widehat{Q}_{\psi}(r_p)  \right)}{\sum_{t=1}^{T}I \left( r_{p,t} \leq \widehat{Q}_{\psi}(r_p)  \right)},
\end{eqnarray}
\begin{equation} \label{PSI2HAT}
\widehat{\Psi}_2(r_p,\psi)=\frac{\sum_{t=1}^{T}r_{p,t} I \left(0 \leq r_{p,t} \leq \widehat{Q}_{\psi}(r_p)  \right)}{\left| \sum_{t=1}^{T}r_{p,t} I \left( r_{p,t} < 0  \right)\right|},
\end{equation}
where $r_{p,t}$ denotes the portfolio return observed at $t$, $\widehat{Q}_{\psi}(r_p)$ denotes the estimated $\psi$-th quantile of the portfolio returns, $I(\cdot)$ is the indicator function taking value 1 if the condition in $(\cdot)$ is true, 0 otherwise. 

Beside highlighting the impact of the quantile regression model in terms of $\Psi_1(R_p,\psi)$ and $\Psi_2(R_p,\psi)$, we go further considering also the central $\vartheta$ values. Now we focus on portfolio volatility, quantified by the mean absolute deviation introduced by \cite{KoYa91}:
\begin{eqnarray} \label{MAD}
MAD(R_p)=\mathbb{E}\left[\left|R_p-\mathbb{E}[R_p]\right|\right],
\end{eqnarray}
estimated in empirical applications as 
\begin{eqnarray} \label{ESTMAD}
\widehat{MAD}(r_p)=\frac{1}{T}\sum_{t=1}^{T} \left| r_{p,t} - \bar{r}_p \right|,
\end{eqnarray}
where $\bar{r}_p$ is the sample mean of portfolio returns in the interval $[1,T]$. 

Under the hypotheses that the portfolio mean, $\mathbb{E}[R_p]$, and the median regression intercept $\xi(\vartheta=0.5)$ are both equal to zero, we show that the median regression allows to minimize the quantity in (\ref{MAD}). Indeed, the quantile regression model at $\vartheta=0.5$ minimizes  $\mathbb{E} [|R_n-w_1(0.5)R^*_1-...-w_{n-1}(0.5)R^*_{n-1}-\xi(0.5)|]=\mathbb{E}[ |R_p-\xi(0.5)|]$. Thus, under the assumptions $\mathbb{E}[R_p]=0$ and $\xi(0.5)=0$, the median regression minimizes the mean absolute deviation of portfolio returns. 

To summarize, the quantile regression model allows reaching different purposes. First of all, we should choose a low probability level, $\alpha$, when we want to minimize the extreme risk, quantified by the $\alpha$-risk. When the attention is focused on volatility minimization, quantified by $MAD$, we should use median regression. Finally, with a high probability level, $\psi$, we minimize $\Psi_1(R_p,\psi)$, with positive effects in terms of $\Psi_2(R_p,\psi)$. 

\subsubsection{Simulation exercise}

\cite{BaKoKo04} applied the model in (\ref{quantmodalprisk}) by using simulated returns of 4 assets, showing its better performance in terms of extreme risk with respect to the classic \cite{Ma52} portfolio. Nevertheless, in the real world, investors trade financial portfolios consisting of many more assets, primarily to achieve a satisfactory diversification level, to better deal with the risk-return trade-off \citep{Ma52}. In order to further motivate the relevance of quantile regression approaches for portfolio allocation, and to show the impact of the methodological improvements previously described, we consider a simulation exercise on portfolios containing 94 assets.\footnote{The portfolios dimensionality comes from the fact that we simulated returns from a distribution whose covariance matrix and mean vector are estimated from real data. We refer here to the constituents of the Standard \& Poor's 100 index at November 21, 2014, and whose time series are continuously available from November 4, 2004 to November 21, 2014. See Section \ref{sec:data_descr} for further details on the dataset.}
The returns are simulated from distributions with different features, in terms of kurtosis and skewness, to verify how these differences impact on the performance of the considered strategies. In particular, we test 4 different distributions: the Multivariate Normal, the Multivariate $t$-Student with 5 degrees of freedom, the Multivariate Skew-Normal with negative skewness and  the Multivariate Skew-Normal with positive skewness. In the case of the Multivariate Skew-Normal, we used two different values of the skewness parameter, to obtain returns series with average skewness equal to 0.02 and -0.02, respectively.

\begin{figure}[htb!]
\centering
\hbox{\hspace{-0.3cm}\includegraphics[scale=0.92]{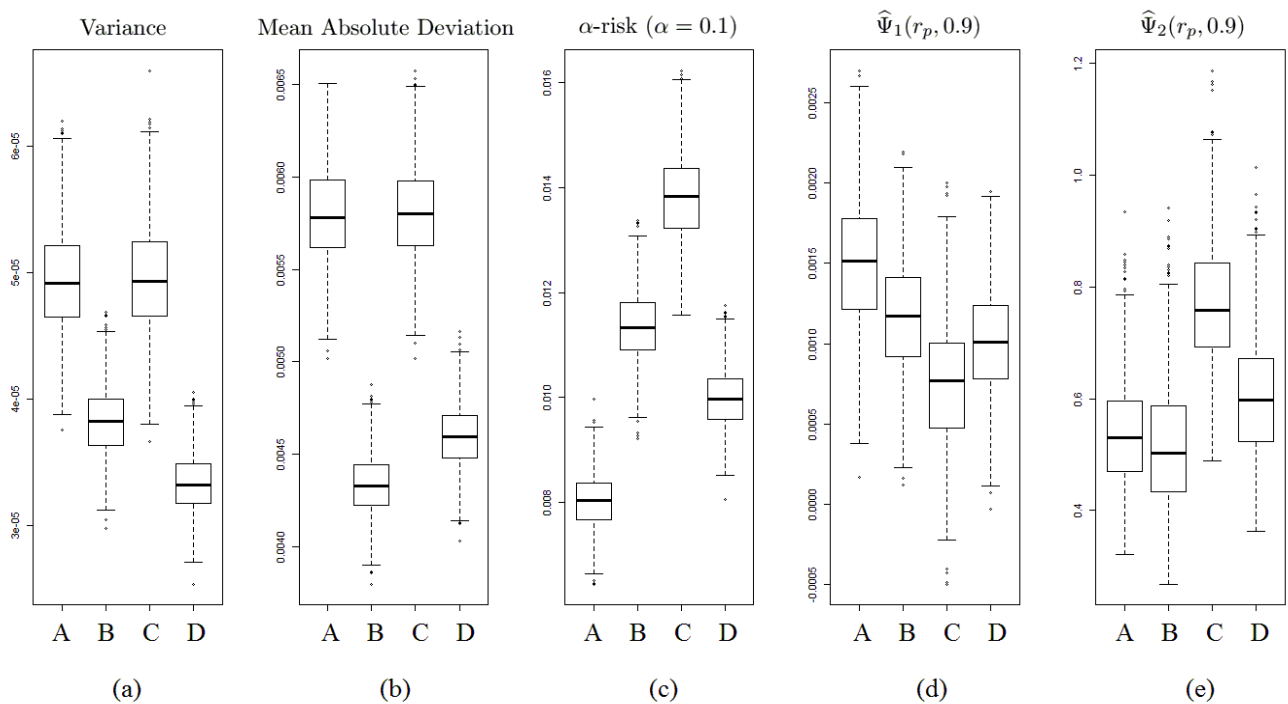}}
\caption{\footnotesize{In-Sample results with returns simulated from Multivariate Normal Distribution, whose covariance matrix and mean vector are estimated from real data, i.e. the constituents of the Standard \& Poor's 100 index at November 21, 2014, and whose time series are continuously available from November 4, 2004 to November 21, 2014. The analysis is carried out on 1000 simulated samples of 94 assets returns for 500 periods. From left to right the subfigures report the boxplots of the following statistics: variance (a), mean absolute deviation (b), $\alpha$-risk at $\alpha=0.1$ (c), $\widehat{\Psi}_1(r_p,\psi)$ at $\psi=0.9$ (d) and $\widehat{\Psi}_2(r_p,\psi)$ at $\psi=0.9$ (e). A, B, C  denote the strategies built from quantile regression models applied at probabilities levels of 0.1, 0.5 and 0.9, respectively, whereas D refers to the ordinary least squares regression model.}}
\label{Boxplot_Normal}
\end{figure}

\begin{figure}[htb!]
\centering
\hbox{\hspace{-0.3cm}\includegraphics[scale=0.92]{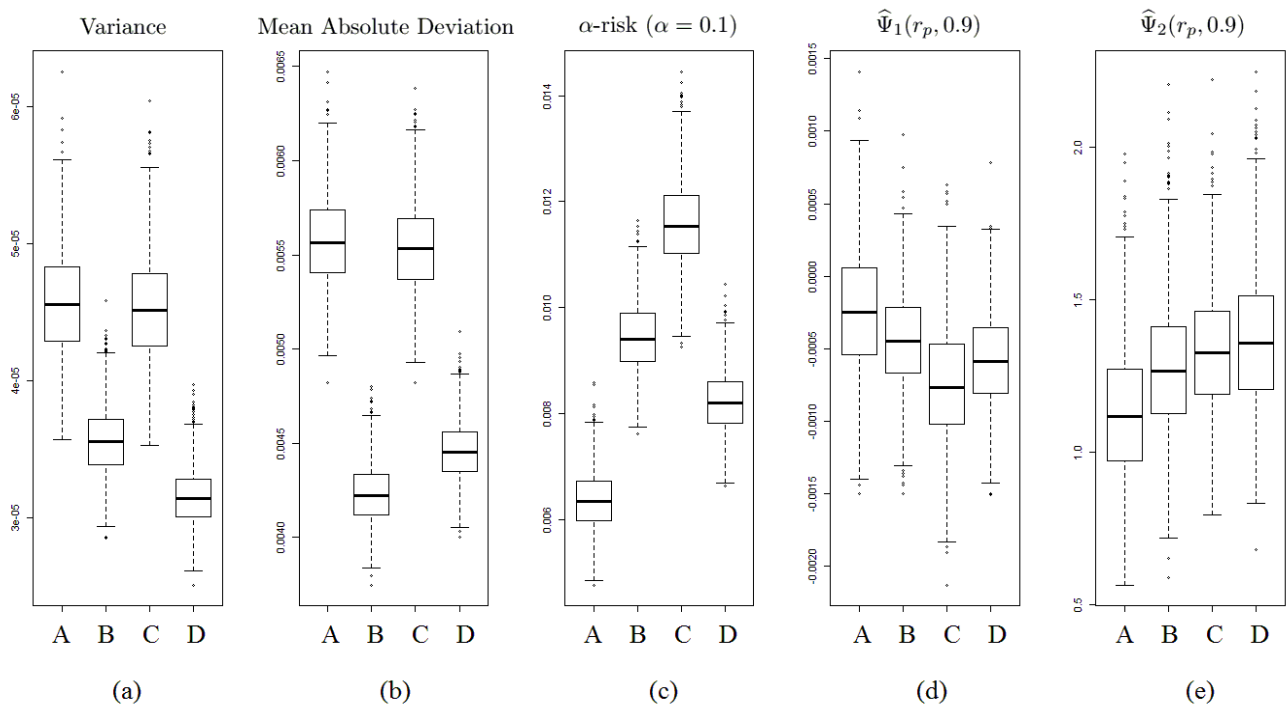}}
\caption{\footnotesize{In-Sample results with returns simulated from Multivariate Skew-Normal Distribution (average skewness of 0.02), whose covariance matrix is estimated from real data, i.e. the constituents of the Standard \& Poor's 100 index at November 21, 2014, and whose time series are continuously available from November 4, 2004 to November 21, 2014. The analysis is carried out on 1000 simulated samples of 94 assets returns for 500 periods. From left to right the subfigures report the boxplots of the following statistics: variance (a), mean absolute deviation (b), $\alpha$-risk at $\alpha=0.1$ (c), $\widehat{\Psi}_1(r_p,\psi)$ at $\psi=0.9$ (d) and $\widehat{\Psi}_2(r_p,\psi)$ at $\psi=0.9$ (e). A, B, C  denote the strategies built from quantile regression models applied at probabilities levels of 0.1, 0.5 and 0.9, respectively, whereas D refers to the ordinary least squares regression model.}}
\label{Boxplot_SkewPos}
\end{figure}

We simulated, from each distribution, 1000 samples of 94 assets returns for 500 periods, comparing four strategies: the standard as in (\ref{fanminvareq}), denoted as $OLS$, and the ones arising from the quantile regression models applied at three  probability levels, that is $\vartheta=\{0.1,0.5,0.9\}$, denoted, respectively, as $QR(0.1)$, $QR(0.5)$ and $QR(0.9)$, respectively. The portfolios weights determined by the four strategies are estimated from each of the 1000 simulated samples and the portfolios returns are computed by means of the in-sample approach.\footnote{See Section \ref{sec:roll_an_sec}
for further details about the in-sample analysis.} Thus, for each strategy and for each sample, we obtain 500 portfolio returns from which we compute the following statistics: variance, mean absolute deviation, $\alpha$-risk (with $\alpha=0.1$), $\Psi_1(R_p,\psi)$ and  $\Psi_2(R_p,\psi)$, at $\psi=0.9$. We display in Figures \ref{Boxplot_Normal}-\ref{Boxplot_SkewPos}  the results obtained from the Multivariate Normal and  Skew-Normal (right-skewed) distributions.\footnote{The boxplots obtained in the case of the other distributions are available on request. Results are qualitatively similar to those obtained from the Multivariate Normal distribution: the presence of fatter tails, as in the Multivariate $t$-Student, or of negative asymmetry, as in the Multivariate Skew-Normal with negative skewness, don't lead to different results.}

As expected, $OLS$ and $QR(0.5)$ provide the best results in terms of portfolio volatility, since the former minimizes the portfolio variance (Subfigure (a)), whereas the latter minimizes the portfolio $MAD$ (Subfigure (b)). $QR(0.1)$ minimizes the $\alpha$-risk at $\alpha=0.1$ (Subfigure (c)), differently, $QR(0.9)$ is the best strategy in terms of profitability. Indeed, as it is possible to see from Subfigure (d), it outperforms the other three strategies in terms of $\widehat{\Psi}_1(r_p,0.9)$, which quantifies the average of portfolio returns which are less or equal to their $90$-th percentile. It is interesting to observe that, when the assets returns distributions are positive skewed (Figure \ref{Boxplot_SkewPos}(d)), $\widehat{\Psi}_1(r_p,0.9)$ takes negative values, mainly in the case of $QR(0.9)$; it means that, on average, the positive returns prevail over the negative ones, even if the most favorable outcomes in the right tail of the distributions are discarded. Finally, $QR(0.9)$ provides the best results in terms of $\widehat{\Psi}_2(r_p,0.9)$ in three out of four cases; the exception occurs in Figure \ref{Boxplot_SkewPos}(e), where the returns are characterized by significant positive skewness, which is typically an unrealistic assumption for financial returns series, typically affected by negative skewness, as discussed in \cite{Co01}. The results also show that the ranking based on $\Psi_1(R_p,\psi)$, as expected, might not coincide to the one based on $\Psi_2(R_p,\psi)$. Indeed, we checked that $QR(0.9)$ always provides, also in the case of the Multivariate Skew-Normal distribution (right-skewed), the highest values of the difference between the numerator and the denominator of the ratio defined in Equation (\ref{PSI2}). Nevertheless, in all the cases apart the one in which the distributions of returns are assumed to be right-skewed (this is hardly the case of financial time series), $QR(0.9)$ turns out to be the best strategy in terms of $\Psi_2(R_p,\psi)$ too.

\subsection{The inclusion of the $\ell_1$-norm penalty} \label{sec:method2}

Large portfolios allow taking advantage of the diversification benefits. Nevertheless, \cite{St87} and recently \cite{FaZhYu12} show that the inclusion of additional assets in the portfolio involves relevant benefits but only up to a certain number of assets. On the other hand, the number of parameters to estimate increases as the portfolio dimensionality grows. As a result, the consequent accumulation of estimation errors becomes a problem that must be carefully addressed. For, instance, \cite{KoDoMa12} defined the estimation error as the price to pay for diversification. Furthermore, when large portfolios are built through regression models, as showed in Section 2.1, the assets returns are typically highly correlated; then, the estimated portfolios weights are  poorly determined and exhibit high variance. 

We propose here a further extension to the \cite{BaKoKo04} approach in order to deal with financial portfolios characterized by large cross-sectional dimension, i.e. by a large number of assets. Our solution builds on penalizations techniques, see e.g. \cite{HaTiFr09}, widely applied in the recent financial literature, see e.g. \cite{DMGaNoUp09}, \cite{FaZhYu12}, \cite{FaPaWi14}, \cite{YeYe14}, \cite{AnBa14}. Among all the possible methods, we make use of the $\ell_1$-norm penalty, useful in the context of variable selection, by which we penalize  the absolute size of the regression coefficients. In the last ten years, it became a widely used tool 
not only in linear regression, but also in quantile regression models, see, e.g. \cite{Ko05}, \cite{BeCh11}, \cite{LiZh08}. 

\cite{HaNa14} used the $\ell_1$-norm penalty in a quantile regression model where the response variable is a core asset, represented by the Standard \& Poor's 500 index, whereas the covariates are hedging satellites, i.e. a set of hedge funds. After setting the quantiles levels according to a precise scheme, the aim is to buy the hedge funds whose coefficients, estimated from the penalized quantile regression model, are different from zero. Therefore, in the work by \cite{HaNa14} the penalized quantile regression is used as a security selection tool, in an index tracking framework. In a second step, by placing the focus on the downside risk, \cite{HaNa14} determine the optimal weights of the funds previously selected by optimizing the objective function given by the Cornish-Fisher Value-at-Risk (CF-VaR). Differently, we use a penalized quantile regression model which allows to solve in just one step both the security selection and the asset allocation problems. The response and the covariates are determined from the assets included in the portfolio, without considering externals variables (such as market indices) with the aim to optimize different performance measures according to different $\vartheta$ levels. 
In particular, given $1 \leq k \leq n$, we propose the asset allocation strategy based on the following model:
\begin{eqnarray} \label{lassoquantregrmodel}
\operatorname*{arg\,min}_{(\textbf{w}_{-k}(\vartheta), \xi(\vartheta) )\in \mathbb{R}^n} \sum_{t=1}^{T} \rho_\vartheta \left(r_{k,t}-
\sum_{j \neq k}^{}w_j(\vartheta)r^*_{j,t}-\xi(\vartheta)\right) + \lambda \sum_{j \neq k}^{}|w_j(\vartheta)|,
\end{eqnarray}
where the parameters
$(\xi(\vartheta),\textbf{w}_{-k}(\vartheta))$ depend on the probability level $\vartheta$, $\textbf{w}_{-k}(\vartheta)$ is the weights vector which does not include $w_k$, that is the weight of the $k$-th asset selected in the (\ref{lassoquantregrmodel}) as numeraire, whereas $\lambda$ is the penalty parameter; the larger $\lambda$, the larger is the portfolio sparsity. 
Thus, by penalizing the sum of the absolute coefficients values, i.e. $\ell_1$-norm, some of the weights might converge to zero. 
Moreover, in a financial perspective, this leads to select portfolios with a fewer active positions, with relevant impact on turnover and transactions costs. 

Unlike Model (\ref{quantmodalprisk}), the solutions of (\ref{lassoquantregrmodel}) depend on the choice of the numeraire $R_k$. Indeed the $R_k$ weight is not penalized, given that it is equal to $w_k(\vartheta)=1-\sum_{j \neq k}^{}w_j(\vartheta)$, to satisfy the budget constraint. Therefore, we need to define a criterion to select the numeraire asset among all the available securities. We propose a naive but intuitive solution: we suggest selecting as numeraire the asset characterized by the lowest in-sample $\widehat{\Psi}_1(R_p,\psi)$ value. 
This choice is due to the fact that the numeraire plays a prominent role with respect to the other selected assets, therefore it must record the best expected performance in terms of a specific indicator.

Another important issue refers to the choice of the optimal $\lambda$ value; we remind that the higher $\lambda$, the more sparse is the portfolio. For this purpose, we follow the approach proposed by \cite{BeCh11}. They considered the problem of dealing with a large number of explanatory variables, with respect to the sample size $T$, where only at most $s \leq n$ regressors have a non-null impact on each conditional quantile of the response variable. In this context, where the ordinary quantile regression estimates are not consistent, they showed that, by penalizing the $\ell_1$-norm of the regressors coefficients, the estimates are uniformly consistent over the compact set $\mathcal{U} \subset (0,1)$. In order to determine the optimal $\lambda$ value, they proposed a data-driven method with optimal asymptotic properties. This method takes into account the correlation among the variables involved in the model and leads to different optimal $\lambda$ values according to the $\vartheta$ level. The penalization parameter is built from the random variable
\begin{equation}
\Lambda = T \operatorname*{sup}_{\vartheta \in \mathcal{U}} \operatorname*{max}_{j \neq k} \left| \frac{1}{T} \sum_{t=1}^{T} \left[ \frac{r_{j,t}^*(\vartheta - \textbf{1}_{ \{e_t \leq \vartheta \}}    )}{\hat{\sigma}_j\sqrt{\vartheta (1-\vartheta)}}  \right]   \right|,
\end{equation}
where $e_1,...,e_T$ are i.i.d. uniform $(0,1)$ random variables independently distributed from the covariates, $\textbf{r}^*$, and $\hat{\sigma}_j^2=T^{-1}\sum_{t=1}^{T}(r_{j,t}^*)^2$. As recommended in \cite{BeCh11}, we simulate the $\Lambda$ values, by running 100000 iterations. Hence, the optimal penalty parameter is computed as
\begin{equation} \label{optimallambdamethodbell}
\lambda^*=\frac{\tau \sqrt{\vartheta (1-\vartheta)}}{T},
\end{equation}
where $\tau=2\widehat{Q}_{0.9}(\Lambda|\textbf{r}^*)$ and  $\widehat{Q}_{0.9}(\Lambda|\textbf{r}^*)$ is the $90$-th percentile of $\Lambda$ conditional on the explanatory variables values. The method given by Equation (\ref{optimallambdamethodbell}) is used in the empirical analysis, discussed in Section \ref{sec:emp_res}, in order to determine the optimal value of the penalization parameter $\lambda$.

\section{Empirical set-up}

\subsection{Data description} \label{sec:data_descr}
 
\begin{figure}[htb!]
\centering
\hbox{\hspace{-0.3cm}\includegraphics[scale=0.92]{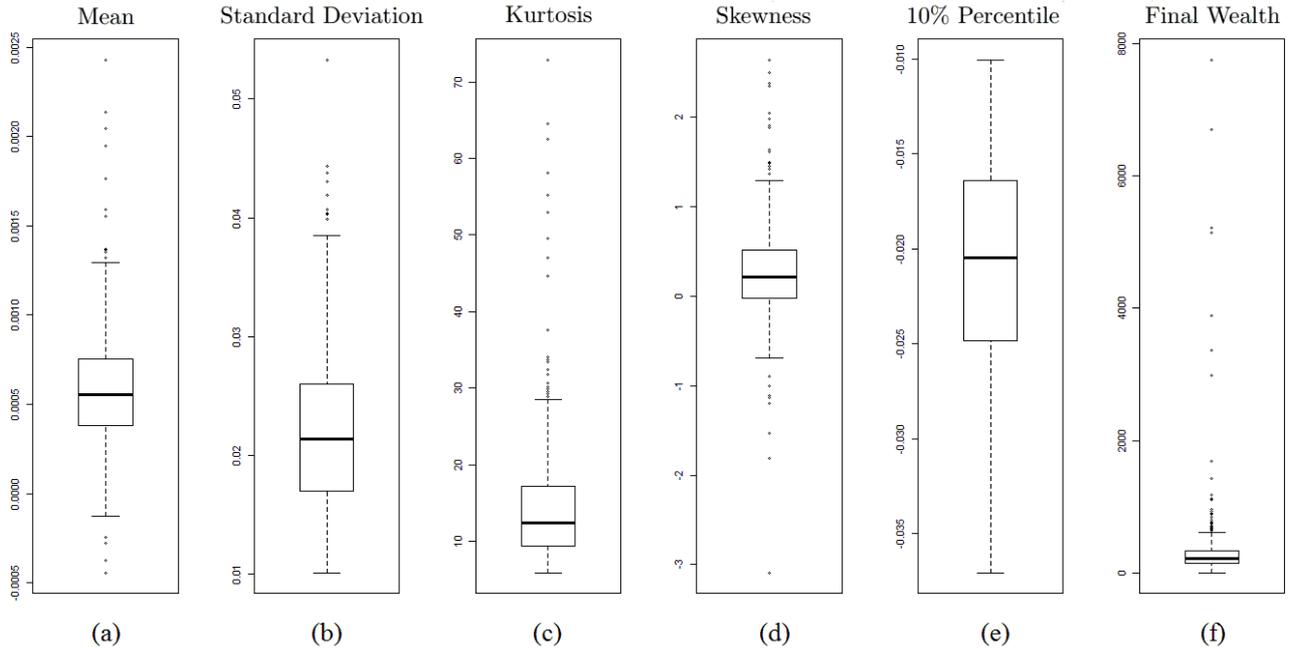}}
\caption{\footnotesize{Statistics computed from each of the assets returns generated by the Standard \& Poor's 500 constituents. From left to right the boxplots refer to: (a) the average returns, (b) the standard deviations of the assets returns, (c) the Kurtosis index, (d) the skewness index, (e) the $10$-th percentile of returns, (f) the final wealth generated by each company at the end of the analysis when the initial wealth is equal to 100 \$.}}
\label{Descr_Stat}
\end{figure}

While \cite{BaKoKo04} used only simulated data in their work, we go further and provide an empirical evaluation based on real data, taking into account two different datasets. They consist of the daily returns of  
the firms included, from November 4, 2004 to November 21, 2014, in the baskets of the Standard \& Poor's 100 and the Standard \& Poor's 500 indices, respectively.\footnote{The data are recovered from Thomson Reuters Datastream.} In the first dataset ($S\&P100$) we deal with 94 assets, whereas in the second one ($S\&P500$) we have 452 stocks. Figure \ref{Descr_Stat} allows to analyze the main descriptive statistics for the largest $S\&P500$ dataset.

From Figure \ref{Descr_Stat}(a) we can see that the average returns are close to zero and that they tend to be positive, being symmetrically distributed around the median $0.055\%$; their maximum and minimum values are equal to $0.242\%$ and $-0.044\%$, respectively. The distribution of the standard deviations is centered at the median value of $2.137\%$ and ranges from $1.013\%$ to $5.329\%$ (Figure \ref{Descr_Stat}(b)), with the presence of a few particularly volatile companies associated to the extreme right-tailed values. The kurtosis index distribution is right-skewed, with  extremely large values in the right tail (Figure \ref{Descr_Stat}(c)): the median is equal to 12.379, whereas the minimum and the maximum are equal, respectively, to 5.948 and 72.867, pointing out that  
the returns distributions are affected by heavy tails, as expected (see \cite{Co01} for stylized facts of financial returns). Figure \ref{Descr_Stat}(d) shows that the skewness index is symmetrically distributed around the median value of 0.209. It ranges from -3.088 to 2.640, with the presence of some extreme values in both the tails; hence, the  returns have leptokurtic and asymmetric distributions. Figure \ref{Descr_Stat}(e) displays the 
$10$-th percentile of the returns series, a measure of extreme risk that we used as estimate of the Value-at-Risk, as discussed in Section
\ref{sec:perf_inds}. From Figure \ref{Descr_Stat}(e) we can see that the distribution of the $10$-th percentile is affected by a slight negative asymmetry, centered at the median of $-2.044\%$, with minimum and maximum equal to $-3.703\%$ and $-1.004\%$, respectively. The last measure, reported in Figure \ref{Descr_Stat}(f), displays the distribution of the wealth we obtain in November 21, 2014 from the single assets, after investing, in November 4, 2004, 100 \$ in each of them. The final wealth distribution, right-skewed, is affected by relevant extreme values in its right tail. It ranges from 2.423 to 7757.073, with median value of 221.650 \$; therefore, on average, we record an increase in stocks values.

\subsection{Rolling analysis with in-sample and out-of-sample assessment} \label{sec:roll_an_sec}

We estimate the portfolios weights by using the least squares and the quantile regression models introduced in Section \ref{sec:method_all}. At the moment in which portfolios are built, it is just possible to predict their future performance, whereas the actual results are not ensured, being evaluable only ex post. Assuming to behave like an investor, we further extend the work of \cite{BaKoKo04} by implementing a rolling window procedure, which allows to analyze the  out-of-sample performance. Furthermore, we can assess the stability of the estimates  over time. 

The rolling window procedure is described as follows. Iteratively, the original assets returns time series, with dimension $T \times n$, are divided in subsamples with window size $ws$. The first subsample includes the daily returns from the first to the $ws$-th day. The second subsample is obtained by removing the oldest observations and including those of the $(ws+1)$-th day.  The procedure goes on till the $(T-1)$-th day is reached. In the empirical analysis we make use of two different window sizes, that is $ws=\{500,1000\}$, to check how the portfolio performance depends on both the portfolio dimensionality and the sample size. 

For each window, we estimate the portfolio weights, denoted by $\widehat{\textbf{w}}_t$, for $t=ws,...,T-1$, by means of a given asset allocation strategy. Let $\textbf{r}_{t-ws+1,t}$ be the $ws \times n$ matrix whose rows are the assets returns vectors recorded in the period between $t-ws+1$ and $t$. Then, portfolio returns are computed both in-sample and out-of-sample. In the first case, for each rolled subsample, we multiply each row of $\textbf{r}_{t-ws+1,t}$ by $\widehat{\textbf{w}}_t$, obtaining $ws$ in-sample portfolio returns, from  
which we can compute the performance indicators described in Section \ref{sec:perf_inds}. 
Overall, from all the $T-ws$ submatrices $\textbf{r}_{t-ws+1,t}$, we obtain $T-ws$ values of each 
performance indicator, whose distribution is analyzed in Section \ref{sec:emp_res}. 

Unlike the in-sample analysis, where we assess estimated performance indicators, the aim of the out-of-sample analysis is to check whether the expectations find confirmation in the actual outcomes. Therefore, the out-of-sample performance plays a critical role, given that it corresponds to the actual impacts on the wealth obtained by an investor that daily revises his portfolio. In particular, for $t=ws,...,T-1$, $\widehat{\textbf{w}}_t$  is multiplied by $\textbf{r}_{t+1}$, i.e. the assets returns vector observed at $t+1$, to obtain the out-of-sample portfolio returns. In this way, for each asset allocation strategy, we obtain one series of out-of-sample portfolio returns, that is a vector with length $T-ws$, from which we compute the performance indicators described in Section \ref{sec:perf_inds}.

\subsection{Performance indicators} \label{sec:perf_inds}

In the empirical analysis we compare several asset allocation strategies, built from both the ordinary least squares and the quantile regression models. We assess and compare their performance using several indicators, to provide information about profitability, risk and impact of trading fees on the portfolios. Each indicator is computed from both the in-sample and out-of-sample returns.

Some of the performance indicators, namely $\alpha$-risk, $MAD$, $\Psi_1(R_p,\psi)$ and $\Psi_2(R_p,\psi)$ are described in Section \ref{sec:method1}. In addition, we take into account other measures, typically used in financial studies. The first one is the Sharpe ratio, a risk-adjusted return measure, defined as 
\begin{eqnarray} \label{SharpeRatio}
SR=\frac{\bar{r}_p}{\hat{\sigma}_p},
\end{eqnarray}
where $\bar{r}_p$ and $\hat{\sigma}_p$ denote, respectively, the sample mean and standard deviation of the portfolio returns.\footnote{To be precise, the numerator of the (\ref{SharpeRatio}) should be equal to the risk-free excess return. For simplicity, we assume that the risk-free return is equal to zero.} As stated in Section \ref{sec:roll_an_sec}, in the in-sample case, $\bar{r}_p$ and $\hat{\sigma}_p$ are computed for each of the rolled subsample $\textbf{r}_{t-ws+1,t}$, for $t=ws,...,T-1$. As result, we obtain $T-ws$ Sharpe ratios. Differently, in the out-of-sample case, we have one portfolio return for each window, obtaining overall a single vector of portfolio returns, with length equal to $T-ws$, from which we compute the Sharpe ratio. 

In addition to the $\alpha$-risk, we also consider the Value-at-Risk, defined as the threshold value such that the probability that the portfolio loss exceeds that value in a given time horizon is equal to $\alpha$. The importance of Value-at-Risk is due not only to the fact that it is widely used by financial institutions to allocate their capital, but it is also used by financial authorities to define capital requirements in their supervisory activity.\footnote{See e.g. "International convergence of capital measurement and capital standards" by Basel Commitee on Banking Supervision, available at 
http://www.bis.org/publ/bcbs118.pdf.}  
In the present work, the Value-at-Risk is estimated as the $\alpha$-th quantile, with $\alpha=0.1$, of the portfolio returns,  both in-sample and out-of-sample. 

Finally, we assess the impact of the trading fees on the portfolio rebalancing activity through the turnover, computed as
\begin{equation} \label{Turnover}
Turn = \frac{1}{T-ws} \sum_{t=ws+1}^{T}\sum_{j=1}^{n} \left| \widehat{w}_{j,t} - \widehat{w}_{j,t-1} \right|,
\end{equation} 
where $\widehat{w}_{j,t}$ is the weight of the $j$-th asset determined by an asset allocation strategy at day $t$. The higher the turnover, the larger is the impact of costs arising from the rebalancing activity.

\section{Empirical results} \label{sec:emp_res}

% LASSO
The first aspect we analyze refers to the impact of the $\ell_1$-norm penalty on portfolio weights. For the quantile regression model, we estimate the optimal shrinkage parameter $\lambda$ according to the method proposed by \cite{BeCh11}. For each quantile level,  we compute the (\ref{optimallambdamethodbell}) by using the full sample data, for both $S\&P100$ and $S\&P500$. Hence, after implementing the rolling window procedure, we compute the number of active and short positions for each rolled sample, whose average values are denoted by $\bar{n}_a$ and $\bar{n}_s$, respectively. The asset allocation strategy built from Model (\ref{lassoquantregrmodel}) is denoted as $PQR(\vartheta)$, for $\vartheta \in (0,1)$. We also apply the $\ell_1$-norm  on Model (\ref{fanminvareq}) and the resulting asset allocation strategy is denoted as $LASSO$. For $LASSO$, $\lambda^*$ is calibrated in order to obtain comparable results, in terms of $\bar{n}_a$, to those generated by the quantile regression model at $\vartheta=0.5$.  For simplicity, we show in Table \ref{lamb_act_short} the $\lambda^*$ values and the average number of active and short positions over the rolled windows just for $PQR(0.1)$, $PQR(0.5)$, $PQR(0.9)$ and $LASSO$.\footnote{In the present work, the position on a certain stock is considered to be long if the 
absolute value of its weight is larger than 0.0005. Similarly, a position is considered short if the weight is less than -0.0005. The weights whose values are between -0.0005 and 0.0005 are set equal to zero.}

\begin{table}[htbp]
\begin{center}
\tiny
\caption{The impact of the $\ell_1$-norm penalty on active and short positions.}
\begin{tabular}{c|c|c|cc|cc|cc|cc}
\hline
Strategy & $\lambda^*$ ($S\&P100$) & $\lambda^*$ ($S\&P500$) & \multicolumn{2}{c|}{$\bar{n}_a$ ($S\&P100$)} & \multicolumn{2}{c|}{$\bar{n}_a$ ($S\&P500$)} & \multicolumn{2}{c|}{$\bar{n}_s$ ($S\&P100$)} & \multicolumn{2}{c}{$\bar{n}_s$ ($S\&P500$)} \\
\hline
   &       &       & $ws$=500 & $ws$=1000 & $ws$=500 & $ws$=1000 & $ws$=500 & $ws$=1000 & $ws$=500 & $ws$=1000 \\
\hline
$PQR(0.1)$ & 0.3644 & 0.6964 & 24    & 43    & 32    & 74    & 6     & 19    & 8     & 26 \\
$PQR(0.5)$ & 0.6073 & 0.1608 & 29    & 44    & 42    & 77    & 8     & 17    & 10    & 28 \\
$PQR(0.9)$ & 0.3644 & 0.6964 & 27    & 41    & 36    & 71    & 7     & 15    & 7     & 23 \\
$LASSO$ & 0.0004 & 0.0004 & 28    & 45    & 42    & 77    & 9     & 20    & 12    & 32 \\
\hline
\end{tabular}\par
\label{lamb_act_short} 
\end{center}
\tiny{The table reports the optimal shrinkage parameters ($\lambda^*$), the average numbers of active ($\bar{n}_a$) and short ($\bar{n}_s$) positions over the rolled samples, for the ordinary least squares ($LASSO$) and the quantile regression ($PQR(\vartheta)$) models, regularized through the $\ell_1$-norm penalty, by using different datasets and window sizes.}
\end{table}

We note that $\lambda^*$ changes according to the $\vartheta$ levels, reaching relatively higher values at the center of the $\vartheta$ domain. This leads to the attenuation of the quantile regression approach tendency for an increase of active positions around the median, see Table \ref{lamb_act_short}. Moreover, we analyzed the evolution over time of the portfolio weights estimated by both the ordinary least squares and the quantile regression approaches. We checked that the weights become more stable with the $\ell_1$-norm penalty and that the effect is more clear with $ws=1000$.
This result is due to the fact that the $\ell_1$-norm penalty shrinks to zero the weights of the highly correlated assets and the larger window size reduces the impact of the estimation errors.

% IN-SAMPLE

\begin{figure}[htb!]
\centering
\hbox{\hspace{-0.3cm}\includegraphics[scale=0.8]{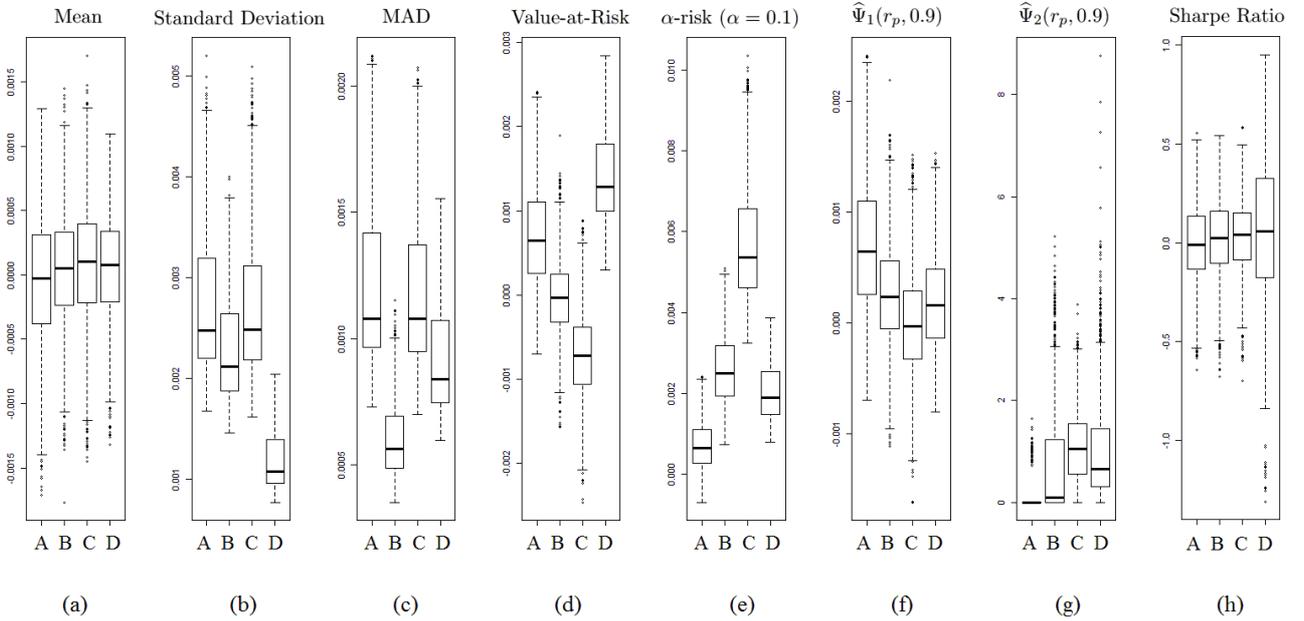}}
\caption{\footnotesize{In-Sample results, obtained without penalizations, from the returns series of the Standard \& Poor's 500 index constituents. Rolling analysis with window size of 500 observations. A, B, C  denote the strategies built from quantile regression models applied at probabilities levels of 0.1, 0.5 and 0.9, respectively, whereas D refers to the ordinary least squares regression model. From left to right the subfigures report the boxplots of the following statistics: mean (a), standard deviation (b), mean absolute deviation (c), Value-at-Risk at the level of 10\% (d), $\alpha$-risk at $\alpha=0.1$ (e), $\widehat{\Psi}_1(r_p,\psi)$ (f), $\widehat{\Psi}_2(r_p,\psi)$ (g), with $\psi=0.9$, Sharpe Ratio (h).}}
\label{IS_sp500}
\end{figure}

\begin{figure}[htb!]
\centering
\hbox{\hspace{-0.3cm}\includegraphics[scale=0.8]{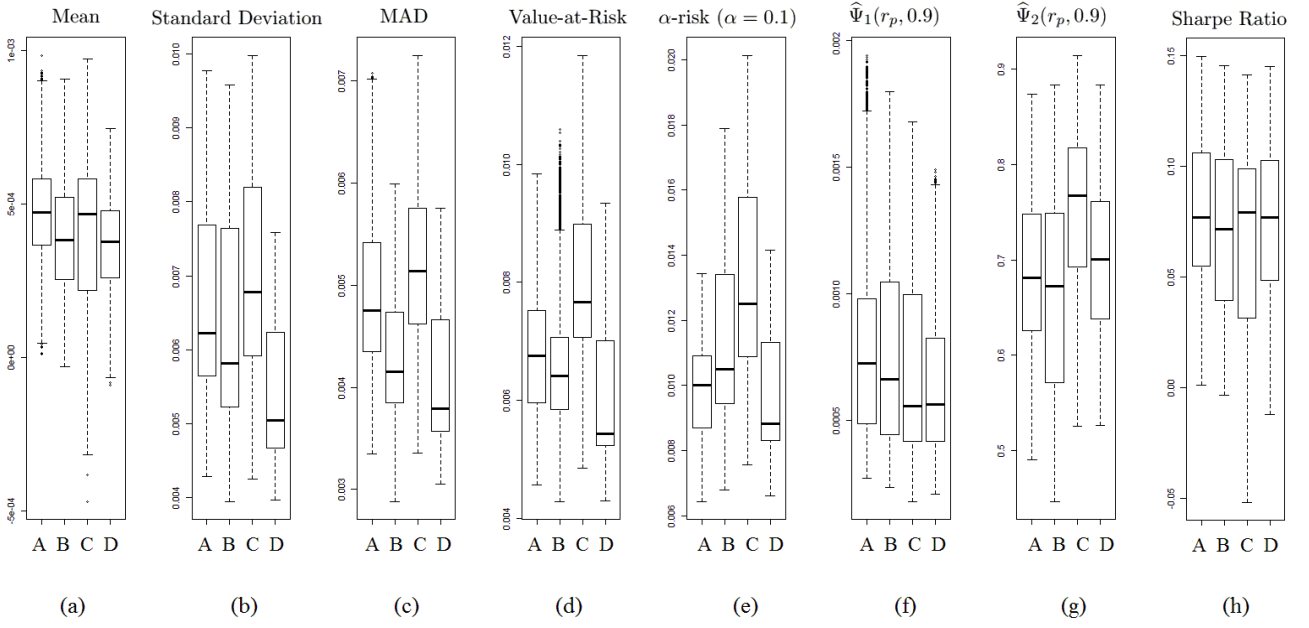}}
\caption{\footnotesize{In-Sample results, obtained by imposing the $\ell_1$-norm, from the returns series of the Standard \& Poor's 500 index constituents. Rolling analysis with window size of 500 observations. A, B, C  denote the strategies built from quantile regression models applied at probabilities levels of 0.1, 0.5 and 0.9, respectively, whereas D refers to the ordinary least squares regression model. From left to right the subfigures report the boxplots of the following statistics: mean (a), standard deviation (b), mean absolute deviation (c), Value-at-Risk at the level of 10\% (d), $\alpha$-risk at $\alpha=0.1$ (e), $\widehat{\Psi}_1(r_p,\psi)$ (f), $\widehat{\Psi}_2(r_p,\psi)$ (g), with $\psi=0.9$, Sharpe Ratio (h).}}
\label{IS_sp500_LASSO}
\end{figure}

Now we analyze and compare the in-sample performance of the various allocation strategies. From the in-sample portfolios returns, computed as described in Section \ref{sec:roll_an_sec}, we obtain the following performance measures: mean, standard deviation, Sharpe Ratio, Value-at-Risk at $\alpha=0.1$ (see Section 3.3), $\alpha$-risk at $\alpha=0.1$, $\Psi_1(R_p,0.9)$, $\Psi_2(R_p,0.9)$ and $MAD$, defined by (\ref{alpharisk}), (\ref{PSI1}), (\ref{PSI2}) and (\ref{MAD}), respectively. As stated in Section \ref{sec:roll_an_sec}, overall, from all the $T-ws$ subsamples arising from the rolling procedure, we get $T-ws$ values of each performance indicator. We make use of boxplots in order to see how the in-sample statistics are distributed over the rolled subsamples. We show in Figures \ref{IS_sp500}-\ref{IS_sp500_LASSO} the output obtained from the ordinary least squares and the quantile regression (applied at $\vartheta=\{0.1,0.5,0.9\}$) models, with and without the imposition of the $\ell_1$-norm penalty, by using as dataset the S\&P 500 constituents ($S\&P500$) and applying the rolling technique with window size of 500 
observations.\footnote{We obtained very similar results, available on request, in the other cases.}

The ordinary least squares model provides the lowest standard deviation, with (Figure \ref{IS_sp500_LASSO}(b)) and without (Figure \ref{IS_sp500}(b)) penalty, as expected, since its objective function is given by the portfolio variance. 
As pointed out in Section \ref{sec:method1}, under the assumptions $\mathbb{E}[R_p]=0$ and $\xi(0.5)=0$, the median regression minimizes the portfolio mean absolute deviation. Such a result is clear with $QR(0.5)$ (Figure \ref{IS_sp500}(c)); when we impose the $\ell_1$-norm penalty, $PQR(0.5)$ still provides the lowest $MAD$ with respect to the other quantile regression models, but it is outperformed by $LASSO$ (Figure \ref{IS_sp500_LASSO}(c)). Indeed, if we impose the $\ell_1$-norm penalty on the median regression, we want to obtain a portfolio characterized by two features: low mean absolute deviation and sparsity, i.e. with a limited number of active positions. Then, the further aim of obtaining sparse portfolios could require sacrifices in terms of higher $MAD$ with respect to the case in which the median regression is not penalized, i.e. when the attention is placed just on the minimization of the quantity defined in Equation (\ref{MAD}). The same phenomenon occurs in the case of $\alpha$-risk: the quantile regression model applied at $\vartheta=0.1$ provides the best results when it is not penalized (Figure \ref{IS_sp500}(e)), but its best performance fades when we impose the $\ell_1$-norm penalty (Figure \ref{IS_sp500_LASSO}(e)). 
In terms of Value-at-Risk there is no strategy which systematically outperforms the other ones; indeed, the ranking based on VaR is not well defined and it varies depending on both the dataset and the window size.
With regard to the Sharpe ratio (Figures \ref{IS_sp500}(h)-\ref{IS_sp500_LASSO}(h)), the lowest variance values allow $OLS$ and $LASSO$ to be ranked on top positions even if they do not generate, on average, remarkable portfolio returns, as showed by Figures \ref{IS_sp500}(a)-\ref{IS_sp500_LASSO}(a). Moreover, when we focus on the average return, there is no strategy systematically dominating the others. Finally, the quantile regression at $\vartheta=0.9$ always outperforms the other strategies in terms of both $\widehat{\Psi}_1(r_p,0.9)$ and $\widehat{\Psi}_2(r_p,0.9)$ (Subfigures (f) and (g)).

\begin{figure}[htb!]
\hbox{\hspace{3cm}\includegraphics[scale=0.6]{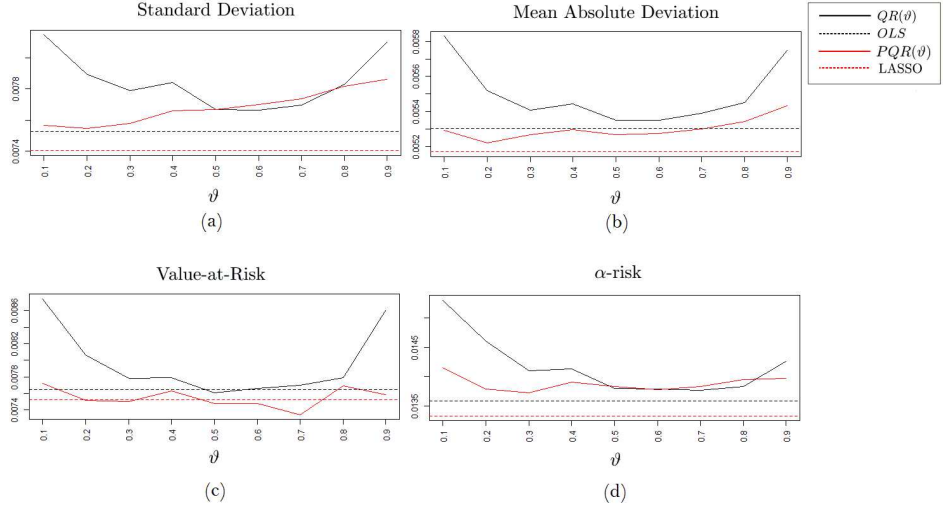}}
\caption{\footnotesize{Out-of-Sample results generated by the strategies built from the ordinary least squares (with ($LASSO$) and without ($OLS$) $\ell_1$-norm penalty) and from the quantile regression (with ($PQR(\vartheta)$) and without ($QR(\vartheta)$) $\ell_1$-norm penalty) models. The strategies are applied on the returns series of the Standard \& Poor's 100 index constituents and the rolling analysis is carried out with a window size of 1000 observations. In the subfigures we consider the following measures: standard deviation (a), mean absolute deviation (b), Value-at-Risk (c) and $\alpha$-risk (d) at the level of 10\%.}}
\label{OOSsp1001}
\end{figure}

\begin{figure}[htb!]
\hbox{\hspace{3cm}\includegraphics[scale=0.6]{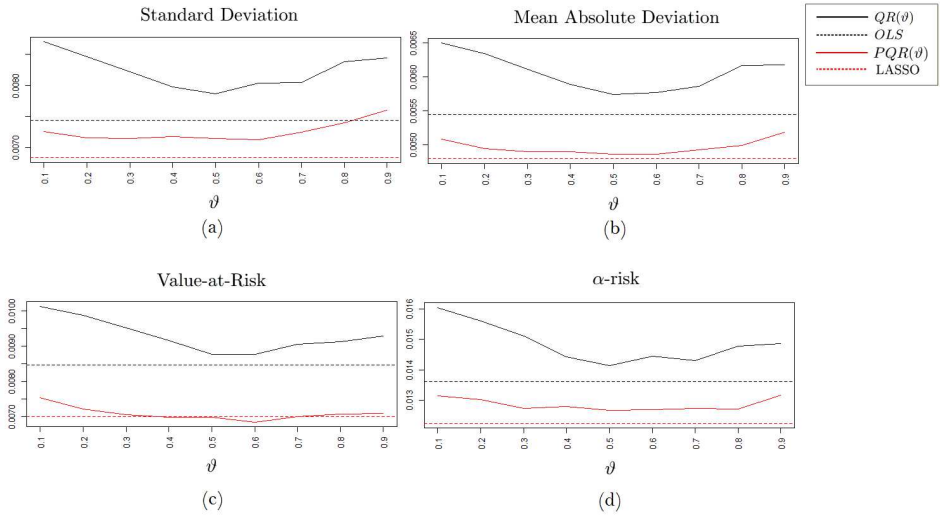}}
\caption{\footnotesize{Out-of-Sample results generated by the strategies built from the ordinary least squares (with ($LASSO$) and without ($OLS$) $\ell_1$-norm penalty) and from the quantile regression (with ($PQR(\vartheta)$) and without ($QR(\vartheta)$) $\ell_1$-norm penalty) models. The strategies are applied on the returns series of the Standard \& Poor's 500 index constituents and the rolling analysis is carried out with a window size of 1000 observations. In the subfigures we consider the following measures: standard deviation (a), mean absolute deviation (b), Value-at-Risk (c) and $\alpha$-risk (d) at the level of 10\%.}}
\label{OOSsp5001}
\end{figure}

\begin{figure}[htb!]
\hbox{\hspace{3cm}\includegraphics[scale=0.6]{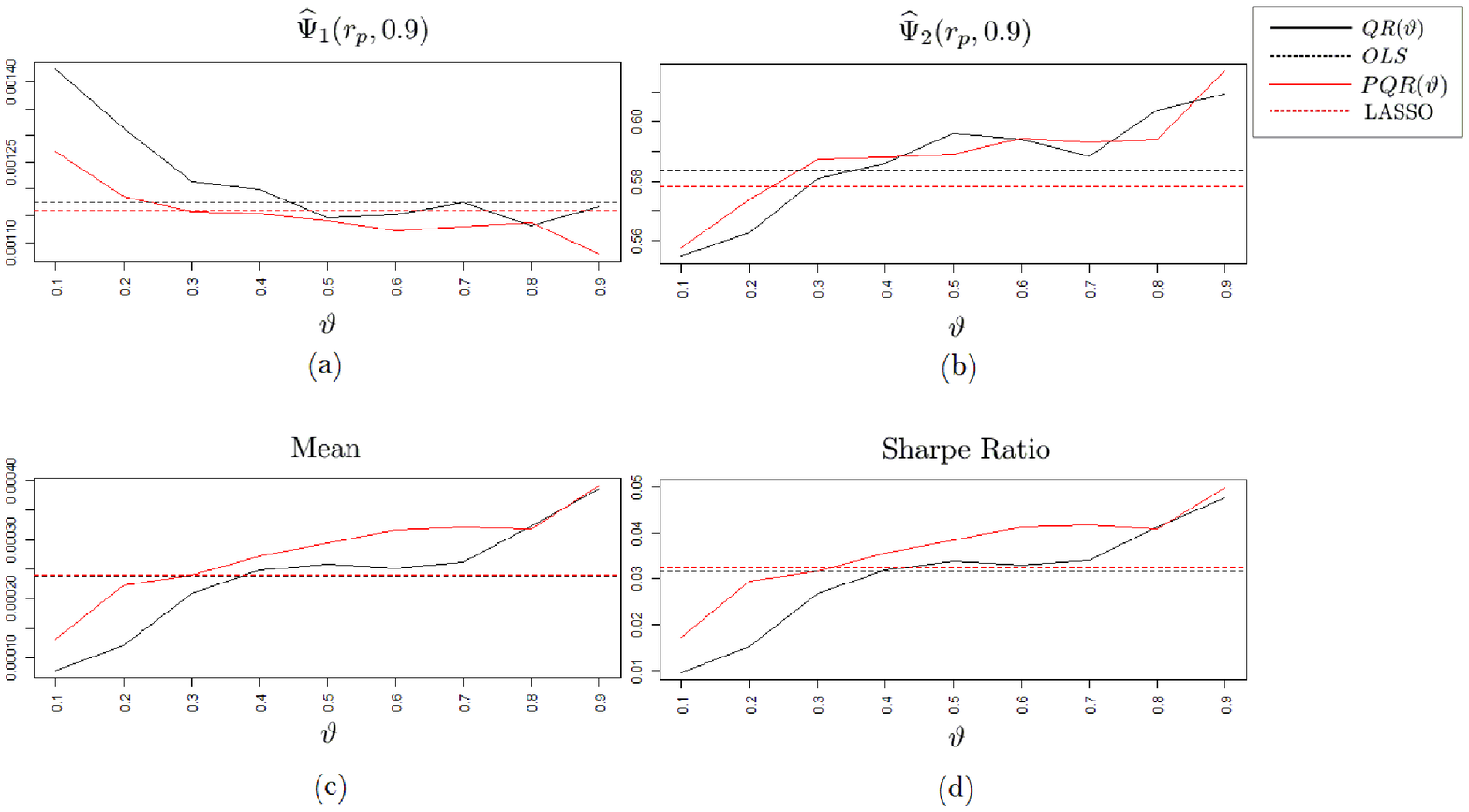}}
\caption{\footnotesize{Out-of-Sample results generated by the strategies built from the ordinary least squares (with ($LASSO$) and without ($OLS$) $\ell_1$-norm penalty) and from the quantile regression (with ($PQR(\vartheta)$) and without ($QR(\vartheta)$) $\ell_1$-norm penalty) models. The strategies are applied on the returns series of the Standard \& Poor's 100 index constituents and the rolling analysis is carried out with a window size of 1000 observations. In the subfigures we consider the following measures: $\widehat{\Psi}_1(r_p,\psi)$ (a) and $\widehat{\Psi}_2(r_p,\psi)$ (b) at $\psi=0.9$, mean (c) and Sharpe Ratio (d).}}
\label{OOSsp1002}
\end{figure}

\begin{figure}[htb!]
\centering
\hbox{\hspace{3cm}\includegraphics[scale=0.6]{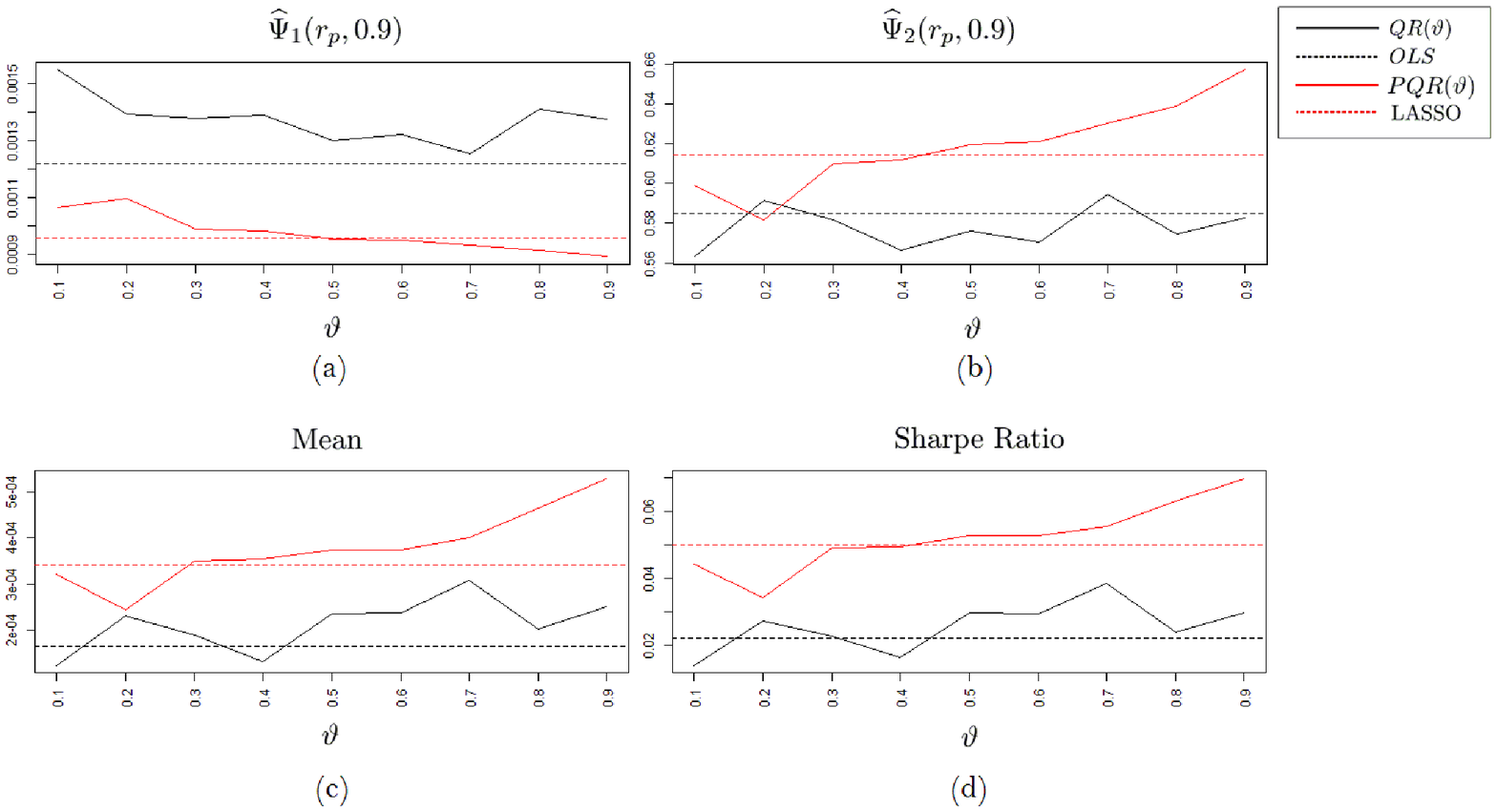}}
\caption{\footnotesize{Out-of-Sample results generated by the strategies built from the ordinary least squares (with ($LASSO$) and without ($OLS$) $\ell_1$-norm penalty) and from the quantile regression (with ($PQR(\vartheta)$) and without ($QR(\vartheta)$) $\ell_1$-norm penalty) models. The strategies are applied on the returns series of the Standard \& Poor's 500 index constituents and the rolling analysis is carried out with a window size of 1000 observations. In the subfigures we consider the following measures: $\widehat{\Psi}_1(r_p,\psi)$ (a) and $\widehat{\Psi}_2(r_p,\psi)$ (b) at $\psi=0.9$, mean (c) and Sharpe Ratio (d).}}
\label{OOSsp5002}
\end{figure}

% OUT-OF-SAMPLE
After analyzing the in-sample results, it is important to check whether they are confirmed out-of-sample. The critical role of the out-of-sample analysis is due to the fact that it refers to the actual performance that an investor 
draws from a financial portfolio. Now we compare the ordinary least squares and the quantile regression (applied at $\vartheta=\{0.1,0.2,0.3,0.4,0.5,0.6,0.7,0.8,$ $0.9\}$) models, with and without $\ell_1$-norm penalty, showing in Figures \ref{OOSsp1001}-\ref{OOSsp5002} just the results obtained by running the rolling method with $ws=1000$; similar results apply for $ws=500$. Consistently to the in-sample expectations, the ordinary least squares regression model records the lowest out-of-sample standard deviation. The $\ell_1$-norm penalty implies important improvements for all the strategies. 
Although the quantile regression model works better in terms of $MAD$ at central $\vartheta$ values, as expected, the ordinary least squares model records the lowest  mean absolute deviations; it is important to notice that the $\ell_1$-norm penalty reduces the $MAD$ of all the strategies (Figures \ref{OOSsp1001}(b)-\ref{OOSsp5001}(b)).

When we analyze the Value-at-Risk, the $\ell_1$-norm penalty brings benefits.
In most cases, the ordinary least squares model outperforms the quantile regression models, mainly when they are not penalized. The $\ell_1$-norm penalty allows to reduce this gap and (see e.g. Figures \ref{OOSsp1001}(c)-\ref{OOSsp5001}(c)) $LASSO$ is outperformed by $PQR(\vartheta)$ at medium-high $\vartheta$ values. The behaviour of the out-of-sample $\alpha$-risk changes with respect to the in-sample case, as we can see, for instance, from Figures \ref{OOSsp1001}(d)-\ref{OOSsp5001}(d). Indeed, the quantile regression model, with and without penalizations, generates disappointing results at low $\vartheta$ levels. The $\ell_1$-norm penalty turns out to be very effective, given that it  reduces the exposure of all the strategies to the $\alpha$-risk. $LASSO$ is ranked as the best strategy and the quantile regression model works better at central $\vartheta$ levels.

\begin{figure}[htb!]
\hbox{\hspace{0cm}\includegraphics[scale=0.9]{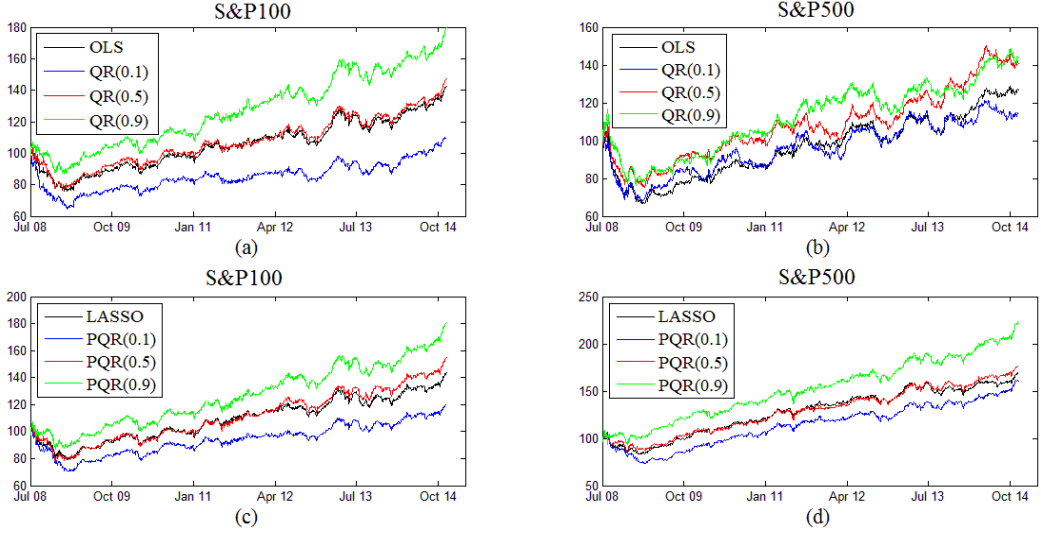}}
\caption{\footnotesize{The evolution of the portfolio value generated by the strategies built from the ordinary least squares (with ($LASSO$) and without ($OLS$) $\ell_1$-norm penalty) and from the quantile regression (with ($PQR(\vartheta)$) and without ($QR(\vartheta)$) $\ell_1$-norm penalty, for $\vartheta=\{0.1,0.5,0.9\}$) models. The subfigures report different results according to the used dataset and to the fact that the regression models are applied with or without $\ell_1$-norm penalty. The rolling analysis is implemented with window size of 1000 observations.}}  
\label{PORTWEAL}
\end{figure}

With regard to profitability, the quantile regression model applied at high $\vartheta$ levels provides outstanding out-of-sample results, consistently to the in-sample expectations. In case the portfolios weights are estimated by using a sample size not sufficiently large with respect to the portfolio dimensionality (i.e. $ws=500$ and $n=452$), $\widehat{\Psi}_1(r_p,0.9)$ and $\widehat{\Psi}_2(r_p,0.9)$ don't exhibit clear trends over $\vartheta$ when quantile regression model is not regularized, as we can see from the  comparisons between Figures \ref{OOSsp1002}(a)-(b) and Figures \ref{OOSsp5002}(a)-(b). Differently, $\widehat{\Psi}_1(r_p,0.9)$ and $\widehat{\Psi}_2(r_p,0.9)$ become, respectively, negative and positive functions of $\vartheta$ 
when we impose the $\ell_1$-norm penalty, also in the case of the $S\&P500$ dataset. Similar conclusions are drawn from the average return and the Sharpe ratio. In particular, in the case of the $S\&P100$ dataset, the quantile regression model applied at high $\vartheta$ levels generates the best performance, with and without penalizations, and $PQR(0.9)$ outperforms all the other strategies (see e.g. Figures \ref{OOSsp1002}(c)-(d)). In the case of the $S\&P 500$ dataset, the quantile regression model at $\vartheta=0.9$ turns out to be the best strategy only if it is applied with the $\ell_1$-norm penalty. In order to analyze the trend over time of the portfolio value generated by each strategy, we assume that the initial wealth is equal to 100 \$ and we update it from $ws+1$ to $T$, according to the out-of-sample portfolio returns. In the case of $S\&P100$, the quantile regression model at $\vartheta=0.9$ not only provides the highest final wealth, but it also dominates the other strategies over time; differently, in the case of the $S\&P500$ dataset, it systematically outperforms the other strategies when 
we impose the $\ell_1$-norm penalty (see Figure \ref{PORTWEAL}).

\begin{figure}[htb!]
\hbox{\hspace{1.5cm}\includegraphics[scale=0.8]{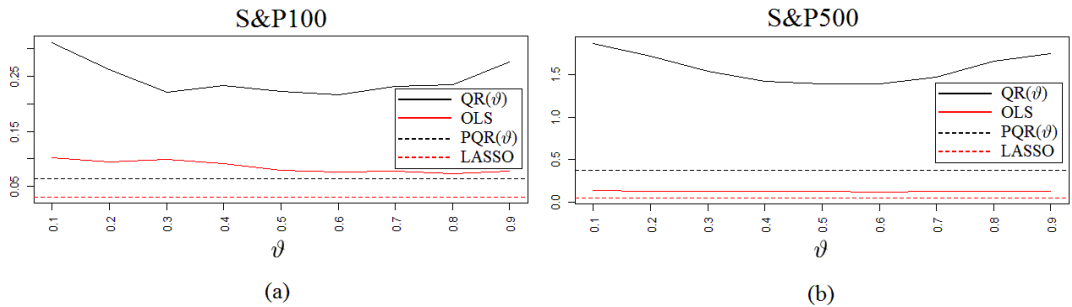}}
\caption{\footnotesize{Turnover of the strategies built from the ordinary least squares (with ($LASSO$) and without ($OLS$) $\ell_1$-norm penalty) and from the quantile regression (with ($PQR(\vartheta)$) and without ($QR(\vartheta)$) $\ell_1$-norm penalty) models. The strategies are applied on the returns series 
of the Standard \& Poor's 100 (Subplot (a)) and 500 (Subplot (b)) indices constituents. The rolling analysis is carried out with window size of 1000 observations.}}
\label{Turnover_pic}
\end{figure}

As stated above, without regularizations, large portfolios are affected by an increasing variability in portfolios weights, with negative effects on trading fees. Consequently, turnover becomes a problem, particularly  for the quantile regression model (Figure \ref{Turnover_pic}). 
The $\ell_1$-norm penalty allows to obtain sparse portfolios, with stable weights over time, thus it turns out to be very effective in terms of turnover control. In fact, the inclusion of the $\ell_1$-norm penalty causes a sharp drop of turnover in all the analyzed cases. The quantile regression model is the one which most benefits from regularization, since the marked gap between $OLS$ and $QR(\vartheta)$ shrinks to become almost irrelevant.

To summarize the out-of-sample results, the $\ell_1$-norm penalty regularizes the portfolio weights, with noticeable positive effects on turnover. Moreover, it leads to clear improvements of both the portfolio risk and profitability. In general, the ordinary least squares model turns out to be the best strategy in terms of risk, given that it implies the lowest levels of volatility and extreme risk; quantile regression model works better at central $\vartheta$ values. The quantile regression, applied at low $\vartheta$ values, generates unsatisfactory results in terms of extreme risk with respect to the in-sample expectations; differently, when we analyze the portfolio profitability and the risk-adjusted return, it provides outstanding performances at high $\vartheta$ levels.

The empirical analysis is applied on data recorded from November 4, 2004 to November 21, 2014. Those years are characterized by special events, namely the subprime crisis, originated in the United States and marked by Lehman Brothers default in September 2008, and the sovereign debt crisis, which hit the Eurozone some years later. Those events had a deep impact on financial markets and, therefore, it is important to check whether they affect the performance of the considered asset allocation strategies. Moreover, in this way, we can analyse whether and how the performance of the strategies depends on the states of the market: the state characterized by financial turmoils and the state of relative calm.    
To this purpose, we divided the series of the out-of-sample portfolios returns into two sub-periods. When the window size is equal to 1000, the first sub-period goes from July 31, 2008 to October 31, 2011, whereas it covers the days between October 31, 2006 and October 31, 2010 at $ws=500$. We can associate this sub-period to the state of financial turmoil, given the proximity to the above mentioned crises. The second sub-period includes the remaining days till November 21, 2014. 

As expected, the strategies record a better out-of-sample performance in the second sub-period with respect to the first one, as we can see, for instance, from  Table \ref{EFFECTCRISIS}, where we report the results obtained from the $S\&P500$ dataset, applying the rolling procedure with window size of 1000 observations.\footnote{The results obtained in the other cases are available on request.}

\begin{center}
\begin{table}[htb]
\begin{center}
\tiny
\captionof{table}{Out-of-Sample Analysis partitioned on two sub-periods.}
\begin{tabular}{ >{\centering\arraybackslash}m{1cm}>{\centering\arraybackslash}m{1.2cm}>{\centering\arraybackslash}m{1.2cm}>{\centering\arraybackslash}m{1.2cm}>{\centering\arraybackslash}m{1.3cm}>{\centering\arraybackslash}m{1.2cm}>{\centering\arraybackslash}m{1.2cm}>{\centering\arraybackslash}m{1.2cm}>{\centering\arraybackslash}m{1.2cm}>{\centering\arraybackslash}m{1.2cm}>{\centering\arraybackslash}m{1.2cm}}
\hline
Strategy & Standard Deviation & $MAD$ & Value at Risk  & $\alpha$-risk  & $\widehat{\Psi}_1(r_p,0.9)$ & $\widehat{\Psi}_2(r_p,0.9)$ & Mean  & Sharpe Ratio & Final Wealth \\
\hline
\multicolumn{10}{c}{FIRST SUB-PERIOD} \\
\hline
$QR(0.1)$	&	1.0041	& 0.7476 &	1.1421	&	1.8869	&	0.1916	&	0.5388	&	0.0004	&	0.0004	&	97.5315	\\
$PQR(0.1)$	&	0.8603	& 0.7476 &	0.8922	&	1.6124	&	0.1417	&	0.5566	&	0.0210	&	0.0244	&	117.1751	\\
$QR(0.5)$	&	0.9228 	& 0.6693 &	1.0251	&	1.6864	&	0.1730	&	0.5295	&	0.0081	&	0.0088	&	106.2985 \\
$PQR(0.5)$	&	0.8668	& 0.6693 &	0.8325	&	1.5876	&	0.1228	&	0.5976	&	0.0350	&	0.0404	&	130.8294	\\
$QR(0.9)$	&	0.9772	& 0.6910 &	1.0271	&	1.7274	&	0.1717	&	0.5393	&	0.0212	&	0.0217	&	116.9810 \\
$PQR(0.9)$	&	0.9167	& 0.6910 &	0.9226	&	1.6384	&	0.1127	&	0.6366	&	0.0557	&	0.0608	&	152.9185	\\
$OLS$	&	0.8771	& 0.6296 &	0.9673	&	1.6514	&	0.1649	&	0.5303	&	-0.0019	&	-0.0022	&	98.0775	\\
$LASSO$	&	0.8140	& 0.6296 &	0.8102	&	1.4955	&	0.1159	&	0.6057	&	0.0360	&	0.0442	&	133.1528	\\
\hline
\multicolumn{10}{c}{SECOND SUB-PERIOD} \\
\hline
$QR(0.1)$	&	0.7135	& 0.5539 &	0.8498	&	1.2700	&	0.1169	&	0.6033	&	0.0238	&	0.0334	&	116.3018	\\
$PQR(0.1)$	&	0.5604	& 0.5539 &	0.6281	&	0.9730	&	0.0702	&	0.6683	&	0.0431	&	0.0768	&	135.6459	\\
$QR(0.5)$	&	0.6215	& 0.4792 &	0.7397	&	1.0842	&	0.0835	&	0.6604	&	0.0389	&	0.0625	&	131.2027	\\
$PQR(0.5)$	&	0.5197	& 0.4792 &	0.5840	&	0.9021	&	0.0639	&	0.6744	&	0.0402	&	0.0773	&	133.4267	\\
$QR(0.9)$	&	0.6875	& 0.5442 &	0.8839	&	1.2280	&	0.0999	&	0.6524	&	0.0292	&	0.0425	&	121.5124	\\
$PQR(0.9)$	&	0.5641	& 0.5442 &	0.6055	&	0.9423	&	0.0634	&	0.6990	&	0.0503	&	0.0892	&	143.6969	\\
$OLS$	&	0.5819	& 0.4584 &	0.7022	&	1.0162	&	0.0766	&	0.6760	&	0.0350	&	0.0602	&	127.9555	\\
$LASSO$	&	0.5227	& 0.4584 &	0.5862	&	0.9094	&	0.0711	&	0.6490	&	0.0324	&	0.0620	&	125.4998	\\
\hline
\end{tabular}\par
\label{EFFECTCRISIS}
\end{center}
\bigskip
\tiny{The table reports the values of different performance indicators generated by the strategies built on the ordinary least squares regression (with ($LASSO$) and without ($OLS$) $\ell_1$-norm penalty) and on  the quantile regression (with ($PQR(\vartheta)$) and without ($QR(\vartheta)$) $\ell_1$-norm penalty, for $\vartheta=\{0.1,0.5,0.9\}$) models. The statistics are computed from 
the Out-of-Sample returns recorded in two different sub-periods: July 2008-October 2011 and October 2011-November 2014. The rolling analysis is applied on the Standard \& Poor's 500 Constituents dataset, with window size of 1000 observations. Standard deviation, $MAD$, Value-at-Risk at $\alpha=0.1$, $\alpha$-risk (with $\alpha=0.1$), $\widehat{\Psi}_1(r_p,0.9)$ and mean are expressed in percentage values. The final wealth is the value we obtain at the end of the analysis from each strategy by investing an initial amount equal to 100 \$.}     
\end{table}
\end{center}

Similarly to the analysis on the entire sample, in both the two sub-periods the ordinary least squares regression almost always records the best performance in terms of risk, evaluated by means of standard deviation, mean absolute deviation, Value-at-Risk and $\alpha$-risk. Differently, the quantile regression model applied at $\vartheta=0.9$ tends to be the best strategy in terms of profitability and risk-adjusted return, as we checked from the average portfolio return, $\widehat{\Psi}_1(r_p,\psi)$, $\widehat{\Psi}_2(r_p,\psi)$, the Sharpe ratio and the final wealth created  after investing an initial amount of 100 \$. 

\subsection{The role of the intercept}

We observed that, in the out-of-sample performance, some strategies, built from quantile regression models at different quantiles levels, kept their in-sample properties, whereas other ones failed in achieving that.
It is important to study the reason underlying this phenomenon and, in the following, we provide one explanation associated with the model intercept. Given the numeraire $R_k$, $1 \leq k \leq n$, and the covariates $R^*_j$, $j \neq k$, for simplicity, we denote the residual term associated with the quantile regression model, applied at the level $\vartheta$, by $\epsilon(\vartheta)$. In the rolling window procedure, the estimated parameters change over time, having their own variability. In order to take into account their dependence on time, we denote by $(\widehat{\xi}_t(\vartheta),\widehat{\textbf{w}}_{-k,t}(\vartheta))$ the coefficients estimated in $t$, from the data recorded in the interval $[t-ws+1;t]$, for $t=ws,...,T-1$. The out-of-sample $\vartheta$-th quantile of $R_k$, computed in $t+1$, depends on both the estimates obtained in $t$ and the realizations of $\textbf{R}$ in $t+1$, being equal to $\widehat{\xi}_t(\vartheta)+\sum_{j \neq k}^{}\widehat{w}_{j,t}(\vartheta)r_{j,t+1}^*$. Therefore the corresponding out-of-sample residual is computed as 
\begin{equation} \label{residual_repre}
\epsilon_{t+1}(\vartheta)=r_{k,t+1}-\left[ \widehat{\xi}_t(\vartheta)+\sum_{j \neq k}^{}\widehat{w}_{j,t}(\vartheta)r_{j,t+1}^*\right].
\end{equation}

Given that the portfolio return, under the budget constraint, can be written as $R_p(\vartheta)=R_k-\sum_{j \neq k}^{}w_{j}(\vartheta)R_{j}^*$, from the (\ref{residual_repre}) we obtain
\begin{equation} \label{portequazz}
r_{p,t+1}(\vartheta)=\epsilon_{t+1}(\vartheta)+\widehat{\xi}_t(\vartheta).
\end{equation}

From the (\ref{portequazz}) we can see that the out-of-sample portfolio return depends on two components: the intercept and the residual. When all the regressors are equal to zero, the estimated intercept corresponds to the estimated quantile of the response variable and, in general, we should expect that $\widehat{\xi}_t(\vartheta)$ is a positive function of $\vartheta$. This phenomenon is particularly accentuated  in case the so-called location-shift hypothesis holds, i.e. when the slopes of the quantile regression models are constant across $\vartheta$, so that the estimated quantiles change according to the intercept levels. Consequently, at high/low $\vartheta$ values, we should expect that the intercept term is a positive/negative component of the portfolio return in (\ref{portequazz}).
Differently, at high/low $\vartheta$ values, the magnitude of the positive residuals is lower/greater than the magnitude of the negative ones; hence, we should expect that $\epsilon_t(\vartheta)$ is a negative/positive component of the portfolio return in the (\ref{portequazz}). 

Given the opposite behaviour of $\epsilon_{t+1}(\vartheta)$ and $\widehat{\xi}_t(\vartheta)$ over $\vartheta$, it is useful to study their distributions 
in order to understand the different out-of-sample performances of the strategies built from the quantile regression models. For simplicity, we compare the 
results obtained from three quantiles levels, i.e. $\vartheta=\{0.1,0.5,0.9\}$. 

We start by analyzing the intercepts distributions, reporting in Table \ref{tab:intercept} the mean and the standard deviation of $\widehat{\xi}_t(\vartheta)$, for $t=ws,...,T-1$.

\begin{table}[htbp]
  \begin{center}
  \tiny
  \caption{Analysis of the intercepts of the quantile regression models.}
\begin{tabular}{ccccc}    
    \hline
          & \multicolumn{2}{c}{$S\&P100$} & \multicolumn{2}{c}{$S\&P500$} \\
    
    Strategy & Mean & St. Dev. & Mean & St. Dev.  \\
    \hline
    \multicolumn{5}{c}{$\vartheta=0.1$} \\
    \hline
    $QR(0.1)$; $ws$=500 & -0.6403 & 0.1943 & -0.0699 & 0.0569 \\
    $PQR(0.1)$; $ws$=500 & -0.7212 & 0.2045 & -0.6910 & 0.1360 \\
    $QR(0.1)$; $ws$=1000 & -0.7482 & 0.0943 & -0.3515 & 0.0371 \\
    $PQR(0.1)$; $ws$=1000 & -0.7464 & 0.0913 & -0.6571 & 0.0754 \\
    \hline
    \multicolumn{5}{c}{$\vartheta=0.5$} \\
    \hline
    $QR(0.5)$; $ws$=500 & 0.0456 & 0.0196 & 0.0026 & 0.0474 \\
    $PQR(0.5)$; $ws$=500 & 0.0601 & 0.0194 & 0.0636 & 0.0220 \\
    $QR(0.5)$; $ws$=1000 & 0.0474 & 0.0171 & 0.0166 & 0.0156 \\
    $PQR(0.5)$; $ws$=1000 & 0.0583 & 0.0120 & 0.0512 & 0.0096 \\
    \hline
    \multicolumn{5}{c}{$\vartheta=0.9$} \\
    \hline
    $QR(0.9)$; $ws$=500 & 0.6680 & 0.1583 & 0.0733 & 0.0520 \\
    $PQR(0.9)$; $ws$=500 & 0.7551 & 0.1204 & 0.7832 & 0.1251 \\
    $QR(0.9)$; $ws$=1000 & 0.7963 & 0.0769 & 0.3780 & 0.0409 \\
    $PQR(0.9$); $ws$=1000 & 0.7908 & 0.0553 & 0.7142 & 0.0626 \\
    \hline
    \end{tabular} \par
  \label{tab:intercept}%
\end{center}
\tiny{The table reports the average values (\%) and the standard deviations (\%) of the intercepts estimated for the quantile regression models with ($PQR(\vartheta)$) and without ($QR(\vartheta)$) $\ell_1$-norm penalty, for $\vartheta=\{0.1,0.5,0.9\}$. The rolling window procedure is applied at $ws=\{500,1000\}$. Datasets: $S\&P100$ and $S\&P500$.} 
\end{table}%

We checked that, as expected, the support of the $\widehat{\xi}(\vartheta)$ distribution moves on the right as $\vartheta$ increases. As a result, from the second and the fourth columns of Table \ref{tab:intercept} it is possible to see that, in average, $\widehat{\xi}_t(\vartheta)$ is a positive component of the out-of-sample portfolio return generated from the quantile regression model at $\vartheta=0.9$; we have the opposite result at $\vartheta=0.1$, whereas at the median level the intercept takes, in average, values close to zero. We can see from the third and the fifth columns of Table \ref{tab:intercept} that, at the median level, the intercept distribution is characterized by the lowest dispersion, consistently to the fact that the median regression implies, among all the quantile regression models, the lowest out-of-sample portfolio volatility. In all the cases, the largest window size of 1000 observations reduces the intercepts dispersions, mainly at $\vartheta=\{0.1,0.9\}$.  

After analyzing the impact of the intercept, we now study the behaviour of the out-of-sample residuals.

\begin{table}[htbp]
 \begin{center}
  \tiny
  \caption{Analysis of the out-of-sample residuals distributions.}
\begin{tabular}{ccccc}    
    \hline
          & \multicolumn{2}{c}{$S\&P100$} & \multicolumn{2}{c}{$S\&P500$} \\
    
    Model & Mean (\%) & St. Dev. (\%) & Mean (\%) & St. Dev. (\%) \\
    \hline
    \multicolumn{5}{c}{$\vartheta=0.1$} \\
    \hline
    $QR(0.1)$; $ws$=500 & 0.6487 & 0.9493 & 0.1455 & 1.8449 \\
    $PQR(0.1)$; $ws$=500 & 0.7407 & 0.8417 & 0.7225 & 0.8339 \\
    $QR(0.1)$; $ws$=1000 & 0.7561 & 0.8217 & 0.3637 & 0.8721 \\
    $PQR(0.1)$; $ws$=1000 & 0.7594 & 0.7637 & 0.6892 & 0.7317 \\
    \hline
    \multicolumn{5}{c}{$\vartheta=0.5$} \\
    \hline
    $QR(0.5)$; $ws$=500 & -0.0249 & 0.8152 & 0.0480 & 1.7475 \\
    $PQR(0.5)$; $ws$=500 & -0.0405 & 0.8169 & -0.0280 & 0.7822 \\
    $QR(0.5)$; $ws$=1000 & -0.0215 & 0.7670 & 0.0069 & 0.7867 \\
    $PQR(0.5)$; $ws$=1000 & -0.0290 & 0.7668 & -0.0137 & 0.7143 \\
    \hline
    \multicolumn{5}{c}{$\vartheta=0.9$} \\
    \hline
    $QR(0.9)$; $ws$=500 & -0.6348 & 0.9144 & -0.0281 & 1.8113 \\
    $PQR(0.9)$; $ws$=500 & -0.7250 & 0.8540 & -0.7398 & 0.8932 \\
    $QR(0.9)$; $ws$=1000 & -0.7578 & 0.8146 & -0.3529 & 0.8446 \\
    $PQR(0.9)$; $ws$=1000 & -0.7517 & 0.7877 & -0.6612 & 0.7631 \\
    \hline
    \end{tabular} \par
  \label{tab:resid}%
\end{center}  
\tiny{The table reports the average values (\%) and the standard deviations (\%) of the out-of-sample residuals arising from the quantile regression models with ($PQR(\vartheta)$) and without ($QR(\vartheta)$) $\ell_1$-norm penalty, for $\vartheta=\{0.1,0.5,0.9\}$. The rolling window procedure is applied at $ws=\{500,1000\}$. Datasets: $S\&P100$ and $S\&P500$.} 
\end{table}%

In contrast to the intercept case, the residuals supports move on the right as $\vartheta$ decreases, as expected. Consequently, as it is possible to see from the second and the fourth columns of Table \ref{tab:resid}, 
the residuals are, in average,  negative/positive components of the portfolios returns in (\ref{portequazz}) at high/low $\vartheta$ levels. 
By comparing Tables \ref{tab:intercept}
and \ref{tab:resid}, it is important to notice that 
the residuals distributions have larger volatilities with respect to the intercepts ones. 

To summarize, if we build an asset allocation strategy from a quantile regression model with high/low $\vartheta$ levels, we can obtain benefits/losses in terms of positive/negative intercept values. Differently, with low/high $\vartheta$ levels, we derive benefits/losses from the residuals. The opposite effects are, in average, balanced.  Nevertheless, the intercepts distributions have a lower dispersion with respect to the residuals distributions. Then, at high $\vartheta$ values, we obtain benefits from a component (the intercept) characterized by greater stability, but, on the other hand, we are penalized by a second component (the residuals) which are more volatile. Differently, when we use quantile regressions models at low quantiles levels, the benefits of positive residuals are more volatile than the losses of negative intercepts. The more stable benefits characterizing the strategies built from the quantile regression models at high $\vartheta$ levels support their better out-of-sample performance.

\section{Concluding remarks}
We have shown how quantile regression based asset allocation corresponds to the minimization of lower tail risk, mean absolute deviation or maximization of a reward measure, depending on the quantile we are looking at. Within such an analyses we introduced a novel performance measure, that is clearly related to specific portfolio return distribution quantiles. In order to cope with the potentially large cross-sectional dimension of portfolio and at the same time to control for estimation error, we combine quantile regression and regularization, based on the $\ell_1$-norm penalty, to estimate portfolio weights. Our empirical evidences, based both on simulations and real data examples, highlights the features and the benefits of our methodological contributions.
The new tools provided (asset allocation strategies, performance measures and penalization approaches) will be of potential interest in several areas including performance evaluation and the design of asset allocation frameworks. 

Further research high on our agenda is the inclusion of different penalty functions, such as the non-convex ones, that have also a direct interpretation as measures of portfolio diversification (e.g. $\ell_q$-norm with $0 \leq q \leq 1$). They not only typically identify investment strategies with  better out of sample portfolio performance, but also promote more sparsity than the  $\ell_1$-norm penalty. Moreover, we aim at developing a method to choose simultaneously the optimal quantile level as well as the optimal intensity of the penalty.

\bigskip

\textbf{Acknowledgments}. The second author acknowledges financial support from the European Union, Seventh Framework Program FP7/2007-2013 under grant agreement SYRTO-SSH-2012-320270, from the MIUR PRIN project MISURA - Multivariate Statistical Models for Risk Assessment, from the Global Risk Institute in Financial Services and from the Louis Bachelier Institute. The third author acknowledges financial support from ICT COST ACTION 1408 - CRONOS.

\bibliographystyle{apa}
\bibliography{MY}

\begin{thebibliography}{}

\bibitem[\protect\astroncite{Acerbi and Tasche}{2002}]{AcTa02}
Acerbi, C. and Tasche, D. (2002).
\newblock Expected shortfall: A natural coherent alternative to value at risk.
\newblock {\em Economic Notes}, 31:379--388.

\bibitem[\protect\astroncite{Alexander and Baptista}{2002}]{AleBap2002}
Alexander, G. and Baptista, A.~M. (2002).
\newblock Economic implications of using a mean-var model for portfolio
  selection: A comparison with mean-variance analysis.
\newblock {\em Journal of Economic Dynamics and Control}, 26:1159--1193.

\bibitem[\protect\astroncite{Ando and Bai}{2014}]{AnBa14}
Ando, T. and Bai, J. (2014).
\newblock Asset pricing with a general multifactor structure.
\newblock {\em Journal of Financial Econometrics, doi: 10.1093/jjfinec/nbu026}.

\bibitem[\protect\astroncite{Artzner et~al.}{1999}]{ArDeEbHe99}
Artzner, P., Delbaen, F., Eber, J., and Heath, D. (1999).
\newblock Coherent measures of risk.
\newblock {\em Mathematical Finance}, 9:203--228.

\bibitem[\protect\astroncite{Basak and Shapiro}{2001}]{BaSh01}
Basak, S. and Shapiro, A. (2001).
\newblock Value-at-risk based risk management: Optimal policies and asset
  prices.
\newblock {\em The Review of Financial Studies}, 14(2):371--405.

\bibitem[\protect\astroncite{Bassett et~al.}{2004}]{BaKoKo04}
Bassett, G., Koenker, R., and Kordas, G. (2004).
\newblock Pessimistic portfolio allocation and choquet expected utility.
\newblock {\em Journal of Financial Econometrics}, 2(4):477--492.

\bibitem[\protect\astroncite{Belloni and Chernozhukov}{2011}]{BeCh11}
Belloni, A. and Chernozhukov, V. (2011).
\newblock L1-penalized quantile regression in high-dimensional sparse models.
\newblock {\em The Annals of Statistics}, 39(1):82--130.

\bibitem[\protect\astroncite{Britten-Jones}{1999}]{BJ99}
Britten-Jones, M. (1999).
\newblock The sampling error in estimates of mean-variance efficient portfolio
  weights.
\newblock {\em Journal of Finance}, 54(2):655--671.

\bibitem[\protect\astroncite{Brodie et~al.}{2009}]{BrDa09}
Brodie, J., Daubechies, I., Mol, C.~D., Giannone, D., and Loris, I. (2009).
\newblock Sparse and stable markowitz portfolios.
\newblock {\em PNAS}, 106(30):12267--12272.

\bibitem[\protect\astroncite{Brodie}{1993}]{Br93}
Brodie, M. (1993).
\newblock Computing efficient frontiers using estimated parameters.
\newblock {\em Annals of Operations Research}, 45(1):21--58.

\bibitem[\protect\astroncite{Caporin et~al.}{2014}]{CaJaLiMa14}
Caporin, M., Jannin, G., Lisi, F., and Maillet, B. (2014).
\newblock A survey on the four families of performance measures.
\newblock {\em The Journal of Economic Surveys}, 28(5):917--942.

\bibitem[\protect\astroncite{Chopra and Ziemba}{1993}]{ChZi93}
Chopra, V.~K. and Ziemba, T. (1993).
\newblock The effect of errors in means, variances and covariances on optimal
  portfolio choice.
\newblock {\em Journal of Portfolio Management}, 19(2):6--11.

\bibitem[\protect\astroncite{Ciliberti et~al.}{2007}]{CiKoMe07}
Ciliberti, S., Kondor, I., and Mezard, M. (2007).
\newblock On the feasibility of portfolio optimization under expected
  shortfall.
\newblock {\em Quantitative Finance}, 7(4):389--396.

\bibitem[\protect\astroncite{Cont}{2001}]{Co01}
Cont, R. (2001).
\newblock Empirical properties of asset returns: stylized facts and statistical
  issues.
\newblock {\em Quantitative Finance}, 1(2):223--236.

\bibitem[\protect\astroncite{DeMiguel et~al.}{2009}]{DMGaNoUp09}
DeMiguel, V., Garlappi, L., Nogales, F.~J., and Uppal, R. (2009).
\newblock A generalized approach to portfolio optimization: Improving
  performance by constraining portfolio norms.
\newblock {\em Management Science}, 55(5):798--812.

\bibitem[\protect\astroncite{Fan et~al.}{2012}]{FaZhYu12}
Fan, J., Zhang, J., and Yu, K. (2012).
\newblock Vast portfolio selection with gross-exposure constraints.
\newblock {\em Journal of the American Statistical Association},
  107(498):592--606.

\bibitem[\protect\astroncite{Farinelli et~al.}{2008}]{FaFeRoThTi2008}
Farinelli, S., Ferreira, M., Rossello, D., Thoeny, M., and Tibiletti, L.
  (2008).
\newblock Beyond sharpe ratio: Optimal asset allocation using different
  performance ratios.
\newblock {\em Journal of Banking and Finance}, 32(10):2057--2063.

\bibitem[\protect\astroncite{Fastrich et~al.}{2014}]{FaPaWi14}
Fastrich, B., Paterlini, S., and Winker, P. (2014).
\newblock Constructing optimal sparse portfolios using regularization methods.
\newblock {\em Computational Management Science, doi
  10.1007/s10287-014-0227-5}.

\bibitem[\protect\astroncite{H\"ardle et~al.}{2014}]{HaNa14}
H\"ardle, W.~K., Nasekin, S., Chuen, D. L.~K., and Fai, P.~K. (2014).
\newblock Tedas - tail event driven asset allocation.
\newblock {\em SFB 649 Discussion Papers SFB649DP2014-032,
  Sonderforschungsbereich 649, Humboldt University, Berlin, Germany}.

\bibitem[\protect\astroncite{Hastie et~al.}{2009}]{HaTiFr09}
Hastie, T., Tibshirani, R., and Friedman, J. (2009).
\newblock {\em The Elements of Statistical Learning}.
\newblock Springer.

\bibitem[\protect\astroncite{Keating and Shadwick}{2002}]{KeSh02}
Keating, C. and Shadwick, W.~F. (2002).
\newblock A universal performance measure.
\newblock {\em The Finance Development Centre, London}.

\bibitem[\protect\astroncite{Koenker}{2005}]{Ko05}
Koenker, R. (2005).
\newblock {\em Quantile regression}.
\newblock Number~38. Cambridge university press.

\bibitem[\protect\astroncite{Koenker and Bassett}{1978}]{KoBa78}
Koenker, R. and Bassett, G. (1978).
\newblock Regression quantiles.
\newblock {\em Econometrica}, 46(1):33--50.

\bibitem[\protect\astroncite{Konno and Yamazaki}{1991}]{KoYa91}
Konno, H. and Yamazaki, H. (1991).
\newblock Mean-absolute deviation portfolio optimization model and its
  applications to tokyo stock market.
\newblock {\em Management Science}, 37(5):519--531.

\bibitem[\protect\astroncite{Kourtis et~al.}{2012}]{KoDoMa12}
Kourtis, A., Dotsis, G., and Markellos, R.~N. (2012).
\newblock Parameter uncertainty in portfolio selection: Shrinking the inverse
  covariance matrix.
\newblock {\em Journal of Banking and Finance}, 36(9):2522--2531.

\bibitem[\protect\astroncite{Krokhmal et~al.}{2002}]{KrPaUr02}
Krokhmal, P., Palmquist, J., and Uryasev, S. (2002).
\newblock Portfolio optimization with conditional value-at-risk objective and
  constraints.
\newblock {\em Journal of Risk}, 4(2):43--68.

\bibitem[\protect\astroncite{Li and Zhu}{2008}]{LiZh08}
Li, Y. and Zhu, J. (2008).
\newblock L1-norm quantile regression.
\newblock {\em Journal of Computational and Graphical Statistics},
  17(1):163--185.

\bibitem[\protect\astroncite{Lintner}{1965a}]{Lintner65b}
Lintner, J. (1965a).
\newblock Security prices, risk and maximal gains from diversification.
\newblock {\em Journal of Finance}, 20:587--615.

\bibitem[\protect\astroncite{Lintner}{1965b}]{Lintner65a}
Lintner, J. (1965b).
\newblock The valuation of risk assets and the selection of risky investments
  in stock portfolios and capital budgets.
\newblock {\em Review of Economics and Statstics}, 47:13--37.

\bibitem[\protect\astroncite{Mansini et~al.}{2007}]{MaOgSp07}
Mansini, R., Ogryczak, W., and Speranza, M. (2007).
\newblock Conditional value at risk and related linear programming models for
  portfolio optimization.
\newblock {\em Annals of Operations Research}, 152(1):227--256.

\bibitem[\protect\astroncite{Markowitz}{1952}]{Ma52}
Markowitz, H. (1952).
\newblock Portfolio selection.
\newblock {\em Journal of Finance}, 7:77--91.

\bibitem[\protect\astroncite{Mossin}{1966}]{Mossin66}
Mossin, J. (1966).
\newblock Equilibrium in a capital asset market.
\newblock {\em Econometrica}, 35:768--783.

\bibitem[\protect\astroncite{Ortobelli et~al.}{2005}]{OrRaStFaBi05}
Ortobelli, S., Stoyanov, S., Fabozzi, F., and Biglova, F. (2005).
\newblock The proper use of risk measures in portfolio theory.
\newblock {\em International Journal of Theoretical and Applied Finance},
  8(8):1107--1133.

\bibitem[\protect\astroncite{Rockafellar and Uryasev}{2000}]{RoUr00}
Rockafellar, R. and Uryasev, S. (2000).
\newblock Optimization of conditional var.
\newblock {\em Journal of Risk}, 2:21--41.

\bibitem[\protect\astroncite{Schmeidler}{1989}]{Sc89}
Schmeidler, D. (1989).
\newblock Subjective probability and expected utility without additivity.
\newblock {\em Econometrica}, 57(3):571--587.

\bibitem[\protect\astroncite{Sharpe}{1964}]{Sharpe64}
Sharpe, W. (1964).
\newblock Capital asset prices: A theory of market equilibrium under conditions
  of risk.
\newblock {\em Journal of Finance}, 19:425--442.

\bibitem[\protect\astroncite{Statman}{1987}]{St87}
Statman, M. (1987).
\newblock How many stocks make a diversified portfolio.
\newblock {\em Journal of Financial and Quantitative Analysis}, 22(3):353--363.

\bibitem[\protect\astroncite{Tibshirani}{1996}]{Ti96}
Tibshirani, R. (1996).
\newblock Regression analysis and selection via the lasso.
\newblock {\em Journal of the Royal Statistical Society, Series B},
  58(1):267--288.

\bibitem[\protect\astroncite{Yen and Yen}{2014}]{YeYe14}
Yen, Y. and Yen, T. (2014).
\newblock Solving norm constrained portfolio optimization via coordinate-wise
  descent algorithms.
\newblock {\em Computational Statistics and Data Analysis}, 76:737--759.

\end{thebibliography}

\end{document}